\documentclass[iop]{emulateapj}
\usepackage{mathrsfs}
\usepackage{url}
\usepackage[normalem]{ulem}
\usepackage{graphicx}
\usepackage{subcaption}

\usepackage{natbib}
\usepackage{color}
\bibpunct{(}{)}{;}{a}{}{,}

\makeatletter
\newenvironment{tablehere}
  {\def\@captype{table}}
  {}
\newenvironment{figurehere}
  {\def\@captype{figure}}
  {}
\makeatother

\newcommand\aproxgt{\mathrel{%
      \rlap{\raise 0.511ex \hbox{$>$}}{\lower 0.511ex \hbox{$\sim$}}}}
\newcommand\aproxlt{\mathrel{%
      \rlap{\raise 0.511ex \hbox{$<$}}{\lower 0.511ex \hbox{$\sim$}}}}

\usepackage{url}
\usepackage{xcolor}
\definecolor{xlinkcolor}{cmyk}{1,1,0,0}
\usepackage[bookmarks=false,         
     pdfnewwindow=true,      
     colorlinks=true,    
     linkcolor=xlinkcolor,     
     citecolor=xlinkcolor,     
     filecolor=xlinkcolor,  
     urlcolor=xlinkcolor,      
final=true
 ]{hyperref}

\def\ir1334{{IRAS\,13349+2438}}
\def\mcg6{{MCG--6-30-15}}

\def\kmps{\ifmmode \rm km~s^{-1} \else $\rm km~s^{-1}$\fi}
\def\psqcm{\ifmmode \rm cm^{-2} \else $\rm cm^{-2}$\fi}
\def\Msun{\ifmmode \rm M_{\odot} \else $\rm M_{\odot}$\fi}
\def\Lsun{\ifmmode \rm L_{\odot} \else $\rm L_{\odot}$\fi}

\newcommand{\qo}{\ifmmode q_{\rm o} \else $q_{\rm o}$\fi}
\newcommand{\Ho}{\ifmmode H_{\rm o} \else $H_{\rm o}$\fi}
\newcommand{\ho}{\ifmmode h_{\rm o} \else $h_{\rm o}$\fi}
\newcommand{\ltsim}{\raisebox{-.5ex}{$\;\stackrel{<}{\sim}\;$}}
\newcommand{\gtsim}{\raisebox{-.5ex}{$\;\stackrel{>}{\sim}\;$}}

\def\fake2{\hphantom{3}}

\shorttitle{A Search for H {\sc i} Ly$\alpha$ Counterparts to Ultra-Fast X-ray Outflows}
\shortauthors{Kriss et al.}

\begin{document}

\slugcomment{Accepted for publication in Apj, 04/16/2018}

\title{A Search for H {\sc i} Lyman $\alpha$ Counterparts to Ultra-Fast X-ray Outflows}

\author{
Gerard~A.~Kriss\altaffilmark{1},
Julia~C.~Lee\altaffilmark{2} and
Ashkbiz Danehkar\altaffilmark{3}
}

\altaffiltext{1}{Space Telescope Science Institute, 3700 San Martin
Drive, Baltimore, MD 21218, USA}
\altaffiltext{2}{Harvard University, John A. Paulson School of Engineering \& Applied Science, 29 Oxford Street, Cambridge, MA 02138, USA}
\altaffiltext{3}{Harvard-Smithsonian Center for Astrophysics, 60 Garden Street, Cambridge, MA 02138, USA}

\begin{abstract} 
Prompted by the \ion{H}{1} Ly$\alpha$ absorption associated with
the X-ray ultra-fast outflow at $-17\,300~\rm km~s^{-1}$
in the quasar PG~1211+143,
we have searched archival UV spectra at the expected locations of 
\ion{H}{1} Ly$\alpha$ absorption for a large sample of
ultra-fast outflows identified in
{\it XMM-Newton} and {\it Suzaku} observations.
Sixteen of the X-ray outflows have predicted \ion{H}{1} Ly$\alpha$ wavelengths
falling within the bandpass of spectra from either the {\it Far Ultraviolet
Spectroscopic Explorer} or the {\it Hubble Space Telescope},
although none of the archival observations were simultaneous with the X-ray
observations in which UFOs were detected.
In our spectra
broad features with full-width at half-maximum of $1000~\rm km~s^{-1}$
have 2-$\sigma$ upper limits on the \ion{H}{1} column density of
generally $\ltsim2\times10^{13}~\rm cm^{-2}$.
Using grids of photoionization models covering a broad range of spectral
energy distributions, we find that producing
\ion{Fe}{26} Ly$\alpha$ X-ray absorption with equivalent widths
$>30$ eV and associated \ion{H}{1} Ly$\alpha$ absorption with
$\rm N_{HI}<2\times10^{13}~cm^{-2}$
requires total absorbing column densities
$\rm N_{H}>5\times10^{22}~cm^{-2}$ and ionization parameters
log $\xi~\gtsim~3.7$.
Nevertheless, a wide range of SEDs would predict observable
\ion{H}{1} Ly$\alpha$ absorption if ionization parameters are only slightly
below peak ionization fractions for \ion{Fe}{25} and \ion{Fe}{26}.
The lack of Ly$\alpha$ features in the archival UV spectra indicates that
either the UFOs have very high ionization parameters,
very hard UV-ionizing spectra,
or that they were not present at the time of the UV spectral observations
due to variability.
\end{abstract}



\section{Introduction}
\label{section:intro}

All contemporary models of galaxy evolution require significant levels of
feedback initiated by active galactic nuclei (AGN) at the galaxy center
\citep{Dubois13, Angles17, Beckmann17, Nelson15}.
The energy deposited by these fast, massive winds impacts the structure of
the host galaxy's interstellar medium and can have profound influence on
star formation through mechanisms that expel gas from the galaxy,
heat the interstellar medium, and prevent the formation of giant molecular
clouds and star-forming regions \citep{Faucher13}.
While much of the feedback that limits galaxy growth can be attributed to star
formation \citep{Hopkins14}, it is AGN feedback that provides the link
between galaxy growth and the supermassive black holes at the centers of
galaxies
\citep{Silk98,King03,Ostriker10,Soker10,Faucher12,Zubovas14,Thompson15}.
Making this link requires coupling much of the binding energy of the
central black hole to the host galaxy.
Outflows must tap into as much as 0.5\% \citep{Hopkins10} to 5\%
\citep{DiMatteo05} of the AGN bolometric luminosity to
generate the correlation between black hole mass
and galaxy bulge properties \citep{Ferrarese00, Gebhardt00}.

The outflows most intensively studied in local AGN, the X-ray warm
absorbers and associated UV absorption features \citep{Crenshaw03}
typically have too little kinetic power to meet the above criteria
\citep{Crenshaw12}. However, the ultra-fast X-ray outflows (UFOs) now being
detected in both local and distant AGN
\citep{Tombesi10,Tombesi14a,Tombesi15,Gofford13,Nardini15,Longinotti15,Parker17}
could provide the necessary energy input.
These UFOs are characterized by velocities in excess of 10\,000 $\rm km~s^{-1}$
and high column densities of gas, often exceeding
$10^{22}$--$10^{23}~\rm cm^{-2}$, which would provide sufficient mass and
kinetic energy to influence the evolution of the host galaxy
\citep{Pounds03,Tombesi10,Tombesi14b,King15,Tombesi17}.

Most UFOs are detected only via a single feature near rest energies of 7 keV,
suggesting that they are Lyman $\alpha$ or He $\alpha$ features of
\ion{Fe}{26} or \ion{Fe}{25}.
This lack of spectral features makes secure identification difficult, as well
as establishing the precise physical conditions in the outflow.
Some lower ionization UFOs do show multiple spectral features affording
spectral diagnostics, in particular the multi-component UFO in
the Seyfert 1 galaxy IRAS 17020+4544 \citep{Longinotti15}
(with velocities in the range of 23\,000--30\,000 $\rm km~s^{-1}$),
and the quasar PG 1211+143 \citep{Danehkar18}, with an outflow velocity
of $-17\,300~\rm km~s^{-1}$.
Even more significantly, the outflow in PG 1211+143 has a UV counterpart,
with a broad ($\sim$1000 $\rm km~s^{-1}$) feature detected in \ion{H}{1}
Ly$\alpha$ at $v_{out} = -16\,980~\rm km~s^{-1}$ \citep{Kriss18}.

Given the high ionization of most UFOs, it is important to determine under
what conditions other spectral diagnostics (such as UV absorption lines)
are visible.
\cite{Fukumura2010b} showed that outflows detected via
\ion{Fe}{25} He$\alpha$ can also produce \ion{C}{4} absorption in the
UV if the X-ray to optical luminosity ratio is low enough, e.g.,
$\alpha_{ox} \sim 2$.
Similarly, \cite{Hagino2017} produced a model for the outflow in the
$z = 3.912$ quasar APM 08279+5255 which has both UV and X-ray absorption.
Again, this quasar has a low X-ray to optical luminosity ratio,
$\alpha_{ox} = 1.7$.
Significantly, PG 1211+143 shows only  \ion{H}{1} Ly$\alpha$ and no other
UV absorption lines due to its higher X-ray to optical luminosity ratio
($\alpha_{ox} = 1.46$) \citep{Kriss18}.

It is important to establish under what conditions that gas driving the bulk of
an X-ray UFO can produce both X-ray and UV absorption features.
If the conditions required for the outflow to be dominated by
\ion{Fe}{25} or \ion{Fe}{26}
preclude strong Ly$\alpha$ (or features due to other UV ions),
then the presence of UV absorption lines may be an important diagnostic of
clumpiness in the gas, where higher-density, lower ionization-parameter gas
is associated with the X-ray outflow.
On the other hand, if any detected UV absorption features are compatible with
physical conditions in a homogeneous wind, then UV spectra provide additional
diagnostics of its physical conditions and kinematics.
Spurred by the detection of the \ion{H}{1} Ly$\alpha$ counterpart to the
UFO in PG 1211+143, we undertook an archival study of the UV spectra of other
X-ray UFOs to search for comparable absorption lines.
Our search revealed one possible new counterpart.
In this paper we describe the sample of X-ray UFOs we examined, their archival
spectra, and a study of photoionization models that can produce both
\ion{Fe}{25} or \ion{Fe}{26} and \ion{H}{1} Ly$\alpha$ features.
We conclude with a discussion of implications for further studies of UFOs.

\section{The X-ray UFO Sample and Archival UV Observations}

Our sample of X-ray UFOs is taken from the comprehensive studies of
\cite{Tombesi10} and \cite{Gofford13}.
These two surveys are the most uniform, statistically based searches for UFOs
in AGN to date. While UFOs have been discovered or suggested in
other individual studies, for making a comparison with other wavebands,
using these uniformly selected objects enables us to draw more objective
conclusions.
Among a set of 101 {\it XMM-Newton} observations of 42
radio-quiet AGN, \cite{Tombesi10} detected 18 potential
\ion{Fe}{25} or \ion{Fe}{26} K$\alpha$ features
with outflow velocities $>10\,000~\rm  km~s^{-1}$.
\cite{Gofford13} identified 17 candidate features.
Of these 35 candidates, 16 have UV spectra that cover the wavelength range
predicted for an \ion{H}{1} Ly$\alpha$ counterpart.
The remaining candidates, especially the Seyfert 2s, are too heavily
reddened to produce detectable UV flux.
Several others have outflow velocities placing the predicted wavelength
 for Ly$\alpha$ outside the spectral range of existing UV observations.
The 16 candidates with available UV spectra are listed in Table \ref{HSTObsTbl}.
Most UFOs listed in Table \ref{HSTObsTbl} are from \cite{Tombesi10};
those from \cite{Gofford13} are indicated by footnotes in column (4).
When tabulating the UFO outflow velocities,
a difference from \cite{Tombesi10}
is that we give the outflow velocity as negative, representing a blue shift.
In total, we examined 39 locations in 36 discrete spectra to search
for counterparts to the X-ray UFOs.

Most of the UV spectra in our sample are from observations using the
{\it Far Ultraviolet Spectroscopic Explorer} (FUSE) \citep{Moos00} since
the high outflow velocities shift the predicted wavelength of Ly$\alpha$ to
wavelengths generally shorter than 1150 \AA.
The remaining spectra are from {\it Hubble Space Telescope} (HST) observations
using either the {\it Space Telescope Imaging Spectrograph} (STIS)
\citep{Woodgate98} or the {\it Cosmic Origins Spectrograph} (COS)
\citep{Green12}.
Many of the UFO candidates have multiple UV spectra available.
Unfortunately, none of the UV spectra were simultaneous with any of the
X-ray observations.
Both of these aspects are significant, as \cite{Kriss18} showed that the
Ly$\alpha$ counterpart to the $-17\,300~\rm  km~s^{-1}$ X-ray UFO is
variable, and is only detected in the observation obtained simultaneously
with the {\it Chandra} observation of PG 1211+143 \citep{Danehkar18}.

\begin{table*}
  \caption[]{X-ray UFO Sample}
\label{HSTObsTbl}
\begin{center}
\begin{tabular}{l l c c c c c c c}
\hline\hline
Object & IAU Name & Redshift & $v_{out}$ & UFO Date & UV Data Set & UV & UV Date& Exposure \\
       &          &  ($z$)   &  $( c )$  &          &      & Instrument & & (s) \\
\hline
1H0419$-$577  & J042600.7$-$571202 & 0.104  & $\rm -0.037^a$ & 2002-09-25 & lb4f33030 & HST/COS/G130M & 2010-06-01 & 3\,750 \\
Ark 120  & J051611.4$-$000859 & 0.03271 & $\rm -0.269^a$ & 2003-08-24 & P1011201 & FUSE & 2000-11-01 & 90\,000 \\
Mrk 79   & J074232.8+494834 & 0.02219 & $\rm -0.091^a$ & 2006-09-30 & P1011701 & FUSE & 1999-11-30 & 7\,381 \\
         &                  & 0.02219 & $\rm -0.091^a$ & 2006-09-30 & P1011702 & FUSE & 2000-01-14 & 11\,734 \\
         &                  & 0.02219 & $\rm -0.091^a$ & 2006-09-30 & P1011703 & FUSE & 2000-02-22 & 12\,499 \\
NGC 4051 & J120309.6+443153 & 0.00233 & $\rm -0.150^a$ & 2002-11-22 & C0190101 & FUSE & 2003-01-18 & 13\,933 \\
         &                  & 0.00233 & $\rm -0.150^a$ & 2002-11-22 & C0190102 & FUSE & 2003-03-19 & 28\,594 \\
NGC 4151 & J121032.6+392421 & 0.00332 & $\rm -0.055^b$ & 2006-12-18 & P1110505 & FUSE & 2000-03-05 & 20\,610 \\
         &                  & 0.00332 & $\rm -0.055^b$ & 2006-12-18 & P2110201 & FUSE & 2000-04-08 & 15\,272 \\
         &                  & 0.00332 & $\rm -0.055^b$ & 2006-12-18 & P2110202 & FUSE & 2002-06-01 &  6\,602 \\
NGC 4151 & J121032.6+392421 & 0.00332 & $\rm -0.105^a$ & 2006-11-29 & P1110505 & FUSE & 2000-03-05 & 20\,610 \\
         &                  & 0.00332 & $\rm -0.105^a$ & 2006-11-29 & P2110201 & FUSE & 2001-04-08 & 15\,272 \\
         &                  & 0.00332 & $\rm -0.105^a$ & 2006-11-29 & P2110202 & FUSE & 2002-06-01 &  6\,602 \\
PG 1211+143 & J121417.7+140313 & 0.0809 & $\rm -0.128^a$  & 2001-01-16 & lcs501010 & HST/COS/G140L & 2015-04-12 & 1\,900 \\
            &                  & 0.0809 & $\rm -0.128^a$  & 2001-01-16 & lcs502010 & HST/COS/G140L & 2015-04-14 & 1\,900 \\
            &                  & 0.0809 & $\rm -0.128^a$  & 2001-01-16 & lcs504010 & HST/COS/G140L & 2015-04-14 & 1\,900 \\
           &                   & 0.0809 & $\rm -0.128^a$  & 2001-01-16 & P1072001 & FUSE & 2000-04-25 & 52\,277 \\
Mrk 766  & J121826.5+294846 & 0.01293 & $\rm -0.044^a$ & 2005-05-25 & o5l502030 & HST/STIS/G140L & 2000-04-11 & 2\,925 \\
Mrk 205  & J122144.2+751838 & 0.07084 & $\rm -0.100^a$ & 2000-05-07 & Q1060203 & FUSE & 1999-12-29 & 38\,800 \\
         &                  & 0.07084 & $\rm -0.100^a$ & 2000-05-07 & S6010801 & FUSE & 2002-02-02 & 18\,900 \\
         &                  & 0.07084 & $\rm -0.100^a$ & 2000-05-07 & D0540101 & FUSE & 2003-11-14 & 19\,618 \\
         &                  & 0.07084 & $\rm -0.100^a$ & 2000-05-07 & D0540102 & FUSE & 2003-11-17 & 107\,518 \\
         &                  & 0.07084 & $\rm -0.100^a$ & 2000-05-07 & D0540103 & FUSE & 2003-06-28 & 18\,068 \\
Mrk 279  & J135303.4+691829 & 0.03045 & $\rm -0.220^b$ & 2009-05-14 & P1080303 & FUSE & 1999-12-28 & 61\,139 \\
         &                  &  0.03045 & $\rm -0.220^b$ & 2009-05-14 & P1080304 & FUSE & 2000-01-11 & 30\,288 \\
         &                  &  0.03045 & $\rm -0.220^b$ & 2009-05-14 & S6010501 & FUSE & 2002-01-28 & 18\,115 \\
         &                  &  0.03045 & $\rm -0.220^b$ & 2009-05-14 & S6010502 & FUSE & 2002-01-29 & 19\,082 \\
         &                  &  0.03045 & $\rm -0.220^b$ & 2009-05-14 & C0900201 & FUSE & 2002-05-18 & 47\,414 \\
         &                  &  0.03045 & $\rm -0.220^b$ & 2009-05-14 & D1540101 & FUSE & 2003-05-14 & 91\,402 \\
         &                  &  0.03045 & $\rm -0.220^b$ & 2009-05-14 & $\rm F3250102^c$ & FUSE & 2005-12-06 & 10\,393 \\
Mrk 841  & J150401.2+102616 & 0.03642 & $\rm -0.034^a$ & 2005-07-17 & lc8y15010 & HST/COS/G130M & 2014-07-06 & 1\,740 \\
Mrk 290  & J153552.4+575409 & 0.02958 & $\rm -0.141^a$ & 2006-05-04 & P1072901 & FUSE & 2000-03-16 & 12\,769 \\
         &                     & 0.02958 & $\rm -0.141^a$ & 2006-05-04 & D0760101 & FUSE & 2000-02-27 & 9\,239 \\
         &                     & 0.02958 & $\rm -0.141^a$ & 2006-05-04 & D0760102 & FUSE & 2004-03-06 & 46\,055 \\
         &                     & 0.02958 & $\rm -0.141^a$ & 2006-05-04 & E0840101 & FUSE & 2004-11-07 & 11\,424 \\
Mrk 509  & J204409.8$-$104325 & 0.0344  & $\rm -0.172^a$ & 2000-10-25 & X0170101 & FUSE & 1999-11-02 & 19\,355 \\
         &                    & 0.0344  & $\rm -0.172^a$ & 2000-10-25 & X0170102 & FUSE & 1999-11-06 & 32\,457 \\
         &                    & 0.0344  & $\rm -0.172^a$ & 2000-10-25 & P1080601 & FUSE & 2000-09-05 & 62\,100 \\
Mrk 509  & J204409.8$-$104325 & 0.0344  & $\rm -0.141^a$ & 2005-10-18 & X0170101 & FUSE & 1999-11-02 & 19\,355 \\
         &                    & 0.0344  & $\rm -0.141^a$ & 2005-10-18 & X0170102 & FUSE & 1999-11-06 & 32\,457 \\
         &                    & 0.0344  & $\rm -0.141^a$ & 2005-10-18 & P1080601 & FUSE & 2000-09-05 & 62\,100 \\
Mrk 509  & J204409.8$-$104325 & 0.0344  & $\rm -0.196^a$ & 2006-04-25 & X0170101 & FUSE & 1999-11-02 & 19\,355 \\
         &                    & 0.0344  & $\rm -0.196^a$ & 2006-04-25 & X0170102 & FUSE & 1999-11-06 & 32\,457 \\
         &                    & 0.0344  & $\rm -0.196^a$ & 2006-04-25 & P1080601 & FUSE & 2000-09-05 & 62\,100 \\
MR2251$-178$ & J225405.9$-$173455 & 0.06609 & $\rm -0.137^b$ & 2009-05-07 & P1111010 & FUSE & 2001-06-20 & 54\,112 \\
\hline
\end{tabular}
\end{center}
{\bf Notes.} (1) Object name.
(2) IAU name (Right Ascension and Declination,  J2000 equinox).
(3) Redshift. (4) Outflow velocity of the X-ray UFO.
(5) Date of the X-ray observation of the UFO.
(6) UV data set name. (7) UV instrument used. (8) Date of the UV observation.
(7) Exposure time for the UV observation.\\
$^{\rm a}$ UFO detected by \cite{Tombesi10}.\\
$^{\rm b}$ UFO detected by \cite{Gofford13}.\\
$^{\rm c}$ Sum of four exposures, F3250102, F3250103, F3250104, and F3250106.\\
\end{table*}

The UV spectra in our study were downloaded from the
{\it Mikulski Archive for Space Telescopes} (MAST)
and can be obtained at
\href{https://doi.org/10.17909/T9JQ2K}{10.17909/T9JQ2K}.
FUSE spectra cover the wavelength range from 912--1187 \AA\ using four separate
mirrors and gratings coated with either LiF or SiC.
The MAST data were processed using the final FUSE pipeline \citep{Dixon07},
but after retrieval, we combined the four separate spectra using a
post-processing pipeline that aligned the wavelength scales of all four spectra,
normalized their fluxes to the channel whose center was associated with the
fine-guidance camera, and combined them into a single spectrum to increase
the signal-to-noise (S/N) ratio. These combined spectra were re-binned to
0.05 \AA\ pixels, approximately half a resolution element.
The COS spectra downloaded from MAST were also processed using
customized techniques to align and merge all central wavelength settings
and FP-POS positions \citep{Kriss11b,DeRosa15}.
The single STIS spectrum (Mrk 766) was used as given by MAST.

\section{Analysis of the Spectra}

For each of the UV spectra in our sample, we examined the wavelength
region surrounding the nominal wavelength for Ly$\alpha$
predicted by the outflow velocity for the candidate UFO.
In most cases, a simple linear fit to the continuum was adequate to describe the
spectrum in these narrow wavelength intervals.
In our analysis we ignored foreground interstellar absorption lines, and
potential foreground intergalactic \ion{H}{1} Ly$\alpha$ lines.
When necessary, we fitted more complicated spectral models including the
broad wings of nearby emission lines (e.g., \ion{O}{6} or Ly$\alpha$), or
other intrinsic emission lines.
For our fits we used the task {\tt specfit} \citep{Kriss94}, and used a
criterion of $\Delta\chi^2=3.84$ from the minimum $\chi^2$
(for one degree of freedom--the EW of the absorption feature)
to establish an upper limit at the 2-$\sigma$ confidence level (95\%)
in the case of non-detections.

Of the 36 spectra we examined, only two show evidence for an absorption
feature at the predicted location for \ion{H}{1} Ly$\alpha$.
Both are separate FUSE observations of Mrk 79.
Although the feature in the FUSE spectra of Mrk 79 is formally significant
at a confidence level $>$99.9\%,
it appears in only one of two FUSE detector
segments, and its reality is tentative.

In Figure \ref{FigFitsa} we show portions of the
UV spectrum for each of the
objects in our sample in the velocity range surrounding the previously
detected X-ray UFO feature.
In most cases in which we do not detect any Ly$\alpha$ counterpart, we only
show the highest S/N UV spectrum for that object or for the UFO candidate.
In some cases, e.g. Mrk 79, in which two spectra show a possible UV
counterpart and another one does not, we show all UV spectra.
For Mrk 841, a complicated case in which the anticipated location of a UFO
counterpart is on the shoulder of strong foreground Milky Way Ly$\alpha$
absorption, we show both the raw spectrum and a normalized model.
Finally, for the high velocity ($v_{out} = -0.128c$) UFO in PG 1211+143,
for which the FUSE and COS spectra differ significantly in
spectral resolution, we also show both spectra.
The best fit linear or powerlaw continuum model in each case is shown as a
green line.
For illustrative purposes, we also show what a 1 \AA\ equivalent width (EW)
absorption line with a full-width-at-half-maximum (FWHM) of
$1000~\rm km~s^{-1}$ would look like when placed at the location predicted by
the UFO outflow velocity.
Such a line would be comparable in appearance to the Ly$\alpha$ counterpart
detected in PG 1211+143 at $v_{out} = -16\,980~\rm km~s^{-1}$ by \cite{Kriss18}.

From our fits we either determined upper limits to an absorption feature
of assumed FWHM=1000 $\rm km~s^{-1}$, or measured the EW of the detected
absorption feature.
Given the EW, we also determined an upper limit on the neutral hydrogen column
density by integrating the Gaussian profile of the absorption line using the
apparent optical depth method of \cite{Savage91}.
Table \ref{tab:limits} gives our measurements for each spectrum.

\begin{table*}
  \caption[]{Measurements of \ion{H}{1} Ly$\alpha$ Absorption Features for X-ray UFOs}
  \label{tab:limits}
\begin{center}
\begin{tabular}{l l c c c c}
\hline\hline
{Object} & Data Set & $\lambda_o$ & EW & FWHM & $\rm log~N_{HI}$ \\
      &             & (\AA)       & (\AA) & ($\rm km~s^{-1}$) & ($\rm log~cm^{-2}$) \\
\hline
1H0419$-$577 & lb4f03030 & 1293.3 & $<0.029$ & 1000 & $<12.73$ \\
Ark 120  & P1011201 & $\phantom{0}952.8$ & $<0.42$ & 1000 & $<13.89$ \\
Mrk 79   & P1011701 & 1133.5 & $1.32$ & $1000 \pm 200$ & $14.39$ \\
         & P1011702 & 1134.3 & $<0.19$ & 1000 & $<13.54$ \\
         & P1011703 & 1134.3 & $1.33$ & $910 \pm 200$ & $14.39$ \\
NGC 4051 & C0190101 & 1047.6 & $<0.11$ & 1000 & $<13.31$ \\
         & C0190102 & 1047.6 & $<0.11$ & 1000 & $<13.31$ \\
NGC 4151 & P1110505 & 1097.7 & $<0.079$ & 1000 & $<13.16$ \\
         & P2110201 & 1097.7 & $<0.079$ & 1000 & $<13.16$ \\
         & P2110202 & 1097.7 & $<0.079$ & 1000 & $<13.16$ \\
NGC 4151 & P1110505 & 1154.4 & $<0.072$ & 1000 & $<13.12$ \\
         & P2110201 & 1154.4 & $<0.16$ & 1000 & $<13.46$ \\
         & P2110202 & 1154.4 & $<0.10$ & 1000 & $<13.27$ \\
PG 1211+143 & P1072001 & 1097.7 & $<0.065$ & 1000 & $<13.08$ \\
           & $\rm lcs502020^a$ & 1097.7 & $<0.53$ & 1000 & $<14.00$ \\
Mrk 766  & o5l502030 & 1178.5 & $<1.00$ & 1000 & $<14.27$ \\
Mrk 205  & Q1060203 & 1077.5 & $<0.28$ & 1000 & $<13.72$ \\
         & S6010801 & 1077.5 & $<0.27$ & 1000 & $<13.70$ \\
         & D0540101 & 1077.5 & $<0.25$ & 1000 & $<13.67$ \\
         & D0540102 & 1077.5 & $<0.11$ & 1000 & $<13.31$ \\
         & D0540103 & 1077.5 & $<0.33$ & 1000 & $<13.79$ \\
Mrk 279  & P1080303 & 1001.6 & $<0.11$ & 1000 & $<13.29$ \\
         & P1080304 & 1001.6 & $<0.12$ & 1000 & $<13.33$ \\
         & C0900201 & 1001.6 & $<0.25$ & 1000 & $<13.67$ \\
         & D1540101 & 1001.6 & $<0.054$ & 1000 & $<13.00$ \\
         & S6010501 & 1001.6 & $<0.58$ & 1000 & $<14.03$ \\
         & S6010502 & 1001.6 & $<0.65$ & 1000 & $<14.08$ \\
         & $\rm F3250102^b$ & 1001.6 & $<0.93$ & 1000 & $<14.55$ \\
Mrk 841  & lc8y15010 & 1217.8 & $<0.054$ & 1000 & $<13.00$ \\
Mrk 290  & P1072901 & 1086.0 & $<0.30$ & 1000 & $<13.74$ \\
         & D0760101 & 1086.0 & $<0.38$ & 1000 & $<13.85$ \\
         & D0760102 & 1086.0 & $<0.14$ & 1000 & $<13.41$ \\
         & E0840101 & 1086.0 & $<0.32$ & 1000 & $<13.75$ \\
Mrk 509  & X0170101 & 1057.0 & $<0.053$ & 1000 & $<12.99$ \\
         & X0170102 & 1057.0 & $<0.046$ & 1000 & $<12.93$ \\
         & P1080601 & 1057.0 & $<0.039$ & 1000 & $<12.86$ \\
Mrk 509  & X0170101 & 1091.1 & $<0.074$ & 1000 & $<13.14$ \\
         & X0170102 & 1091.1 & $\ldots$ & 1000 & $\ldots$ \\
         & P1080601 & 1091.1 & $<0.057$ & 1000 & $<13.02$ \\
Mrk 509  & X0170101 & 1031.0 & $<0.059$ & 1000 & $<14.13$ \\
         & X0170102 & 1031.0 & $<0.051$ & 1000 & $<14.13$ \\
         & P1080601 & 1031.0 & $<0.049$ & 1000 & $<14.13$ \\
MR2251$-$178 & P1111010 & 1129.1 & $<0.059$ & 1000 & $<14.13$ \\
\hline
\end{tabular}
\end{center}
$\rm ^a$ Based on sum of all three exposures, lcs501010, lcs502010, and lcs504010.\\
$\rm ^b$ Based on sum of four exposures, F3250102, F3250103, F3250104, and F3250106.\\
\end{table*}
\vspace{12pt}

\section{Discussion of Individual Objects}

Each of the sources and the spectra shown in Figure \ref{FigFitsa}
has its own peculiarities. Below we give additional details on
the UV spectral properties of the sources and our fits to the archival spectra.

\subsection{1H0419$-$577}

The narrow-line Seyfert 1
1H0419$-$577 has fairly strong soft X-ray absorption which
\cite{Pounds04} attribute to low-ionization gas internal to the source.
\cite{Tombesi13} characterize this as a warm absorber whose
parameters are not well determined.
They quote an ionization parameter for this gas of log $\xi \sim 1.2$.
This is roughly compatible with the strong, intrinsic UV absorption
visible in FUSE and HST spectra. \cite{Edmonds11} describe three
low-velocity outflow systems ranging from $-50$ to $-230$ $\rm km~s^{-1}$.
The strongest component, at $-230$ $\rm km~s^{-1}$,
has low density $< 10^3~\rm cm^{-3}$,
and is located at 9 kpc, suggesting it is part of a full-scale
galactic outflow.
Their total spectrum is used in our analysis here.

The two deep, narrow absorption lines at 1286 \AA\ and 1295 \AA\ are
unidentified. They are not foreground interstellar features, but we
consider them to be likely intervening intergalactic (IGM)
absorption. Their equivalent widths of 0.28 \AA\ and 0.13 \AA,
respectively, and their widths, with Doppler parameters
$b = 33$ and $b = 44$ $\rm km~s^{-1}$, are typical of
\ion{H}{1} IGM lines \citep{Tilton12}.

\subsection{Ark 120}

The high velocity of the UFO in Ark 120 places its potential Ly$\alpha$
counterpart near the shortest wavelengths observable with FUSE.
Although this spectrum appears to be of poor quality, most of the
``noise" one sees in Figure \ref{fig:1b} is due to the plethora of foreground
interstellar absorption features present at short UV wavelengths.
These are marked in red in the figure.
Strong Ly$\delta$ and \ion{O}{1} airglow are also present at
948 \AA\ and 950 \AA.
These features and the interstellar absorption lines make it difficult to set
a sensitive upper limit on Ly$\alpha$ absorption in this spectrum.

\subsection{Mrk 79}

This Seyfert 1 has three FUSE observations in the archive. No UV absorption is
detectable in the second observation, which has the highest S/N,
but the first and third data sets show a broad,
shallow depression at the location predicted for the X-ray UFO.
Unfortunately, the center of that feature is coincident with the
\ion{N}{1} $\lambda 1134$ airglow feature,
which is quite prominent in the second observation.
Also, thermal misalignments during the third observation led to
very low fluxes in the LiF2 channel, so the absorption detected mostly in the
LiF1 spectrum cannot be independently confirmed in the LiF2 spectrum.
Although the feature is statistically significant (at a confidence level
exceeding 99.9\%), we consider its detection tentative.

As with 1H0419$-$577, Mrk 79 shows strong, variable
soft X-ray absorption \citep{Gallo11}
due to a warm absorber.
In their analysis of the {\it XMM-Newton} {\tt pn} spectra, \cite{Gallo11} find
significant variations in the ionization parameter,
which is low (log $\xi \sim 0.8-2.0$),
and possible variations in the column density.
Analysis of the longest {\it XMM-Newton} RGS spectrum by \cite{Laha14}
similarly finds log $\xi \sim 2.0$ and log $\rm N_H \sim 21.0~cm^{-2}$.
Mrk 79 also has several intrinsic, variable
UV absorption features due to \ion{O}{6} and Ly$\beta$ in its
FUSE spectrum \citep{Dunn08}.
These have outflow velocities of $-320~\rm and~-1400~km~s^{-1}$,
both of which are at lower velocity than the $< -2400 ~\rm km~s^{-1}$
velocity of the warm absorber in the analysis of the 2008
{\it XMM-Newton} RGS spectrum by \cite{Laha14}.

\subsection{NGC 4051}

Fortuitously, the predicted location of Ly$\alpha$ in this FUSE
spectrum of NGC 4051 lies longward of the wing of \ion{O}{6}
emission in this AGN.
Narrow, foreground interstellar absorption lines are also present,
and they are marked in red in the figure.

\subsection{NGC 4151}

NGC 4151 contains two suggested UFOs, one at $v_{out} = -0.105 c$
\citep{Tombesi10}, and one at $v_{out} = -0.0555 c$
\citep{Gofford13}. Both have predicted locations for
\ion{H}{1} Ly$\alpha$ lying in the FUSE band, for which
three separate observations are available.
The wavelength regions spanned by the predicted wavelengths of
a Ly$\alpha$ counterpart contain several narrow absorption
features that correspond to foreground interstellar lines,
as well as a strong intrinsic absorption line from the
metastable levels of \ion{C}{3}* $\lambda$1176.
Although there are some large-scale undulations in
flux in each of these regions, they are not present in
all observations nor in both detector segments, nor do
they rise to the level of significance of comparable
features seen in the FUSE spectra of Mrk 79.

\subsection{PG 1211+143}

The prototype for this survey, PG 1211+143 has one UFO
at a velocity of $v_{out} = -17\,300~\rm km~s^{-1}$
\citep{Danehkar18} which is associated with \ion{H}{1}
Ly$\alpha$ absorption at $v_{out} = -16\,980~\rm km~s^{-1}$
\citep{Kriss18}.
In the \cite{Tombesi10} sample, PG 1211+143, also has
another UFO at $v_{out} = -0.128 c$, which has been extensively
studied \citep{Pounds14,Pounds16}.
We show two spectra of the region of the predicted Ly$\alpha$
counterpart to this higher velocity UFO in Figures \ref{fig:1i} and \ref{fig:1j}.
The illustrated region is spectrally complex, containing weak
emission lines of \ion{S}{4} and \ion{He}{2} intrinsic to
PG 1211+143, as well as foreground interstellar absorption
features, which are marked in red in the COS spectrum in
Figure \ref{fig:1i}. The COS spectrum was taken with grating G140L,
which has significantly lower resolution (resolving power
$\sim 2000$) than the FUSE spectrum in Figure \ref{fig:1j}
(resolving power $\sim18\,000$), and it is the sum of the three exposures
obtained in the combined {\it Chandra} and {\it HST} campaign
\citep{Danehkar18,Kriss18}.
Neither the FUSE nor the COS spectrum (or the individual exposures in the COS
spectrum) shows evidence for broad Ly$\alpha$ absorption.

\subsection{Mrk 766}

In \cite{Tombesi10}, Mrk 766 has two potential UFOs.
The highest-velocity one falls at wavelengths accessible to
FUSE, but no FUSE observations of Mrk 766 were performed
due to its reddened, faint UV spectrum.
The STIS G140L spectrum shown in Figure \ref{fig:1k} doesn't set a
significantly tight constraint on a potential Ly$\alpha$
feature, both due its low S/N and the low resolution.

\subsection{Mrk 205}

Observation D0540102 of Mrk 205 is representative of its
five separate FUSE observations, none of which show evidence
for any Ly$\alpha$ absorption associated with an X-ray UFO.

\subsection{Mrk 279}

The spectrum shown in Figure \ref{fig:1m} is from the FUSE observations
in the extensive UV and X-ray campaign on
Mrk 279 \citep{Gabel05a,Arav07,Costantini07}.
Mrk 279 has a complex, low velocity outflow, with UV and X-ray
spectral lines spanning velocities from $+100$ to $-550~\rm km~s^{-1}$.
Unfortunately, the broad, blended troughs of intrinsic absorption due to
Ly$\gamma$ and \ion{C}{3} $\lambda$977 in this outflow
fall within the region of interest for
a Ly$\alpha$ counterpart to the suggested X-ray UFO.
This makes it difficult to establish a sensitive upper limit for
any Ly$\alpha$ absorption.

\subsection{Mrk 841}

Unfortunately, the predicted location of Ly$\alpha$ in this spectrum
falls right on the red wing of the damped Ly$\alpha$ absorption due
to foreground neutral hydrogen in the Milky Way.
Nevertheless, we have modeled its profile, the surrounding
AGN continuum, and the blue wing of the broad Ly$\alpha$ emission line.
This model is shown overlayed on the high-quality COS spectrum
in Figure \ref{fig:1n}.
The blue curve in the same panel shows the appearance of the model
if a 1 \AA\ equivalent width absorption line with
FWHM=1000 $\rm km~s^{-1}$ is added. It is clear that this deviates
significantly from the data.
To illustrate this more clearly, Figure \ref{fig:1o} shows a normalized
spectrum where the deep, broad absorption feature is obviously
not a good match to the data.
The red line shows the 2-$\sigma$ upper limit feature that can be
accommodated by our fit.

\subsection{Mrk 290}

Mrk 290 is a typical example of an AGN with a spectral break
in the far UV that extrapolates directly to the
soft X-ray \citep{Shang05}. It also has a complex warm absorber
in the X-ray \citep{Zhang11} with corresponding UV absorption
lines \citep{Zhang15} visible in the FUSE spectrum.
The predicted location of Ly$\alpha$ falls near the gap
in the FUSE LiF1 detector, which is why the S/N drops
abruptly at 1085 \AA.

\subsection{Mrk 509}

As with many objects in this sample, Mrk 509 has a complex
set of intrinsic UV absorption lines associated with a
corresponding set of X-ray warm absorbers.
The UV absorption spans a velocity range of $-400~\rm km~s^{-1}$
to  $+200~\rm km~s^{-1}$ \citep{Kriss00,Kriss11b},
similar to the soft X-ray absorption seen in the X-ray
spectrum \citep{Detmers11,Kaastra11c}.
All the deep, narrow absorption features in Figures \ref{fig:1q}--\ref{fig:1s}
are associated either with the intrinsic absorption features
in Mrk~509 or foreground interstellar lines.

\subsection{MR2251$-$178}

The predicted location of Ly$\alpha$ in this FUSE spectrum of
MR2251$-$178 lies on the far wing of broad \ion{O}{6} emission
intrinsic to MR2251$-$178.
The narrow absorption lines in the spectrum are foreground
interstellar lines.
Prominent \ion{N}{1} airglow is present at $\lambda$1134.

\begin{figure}[t]
\centering
\includegraphics[angle=-90, width=0.52\textwidth]{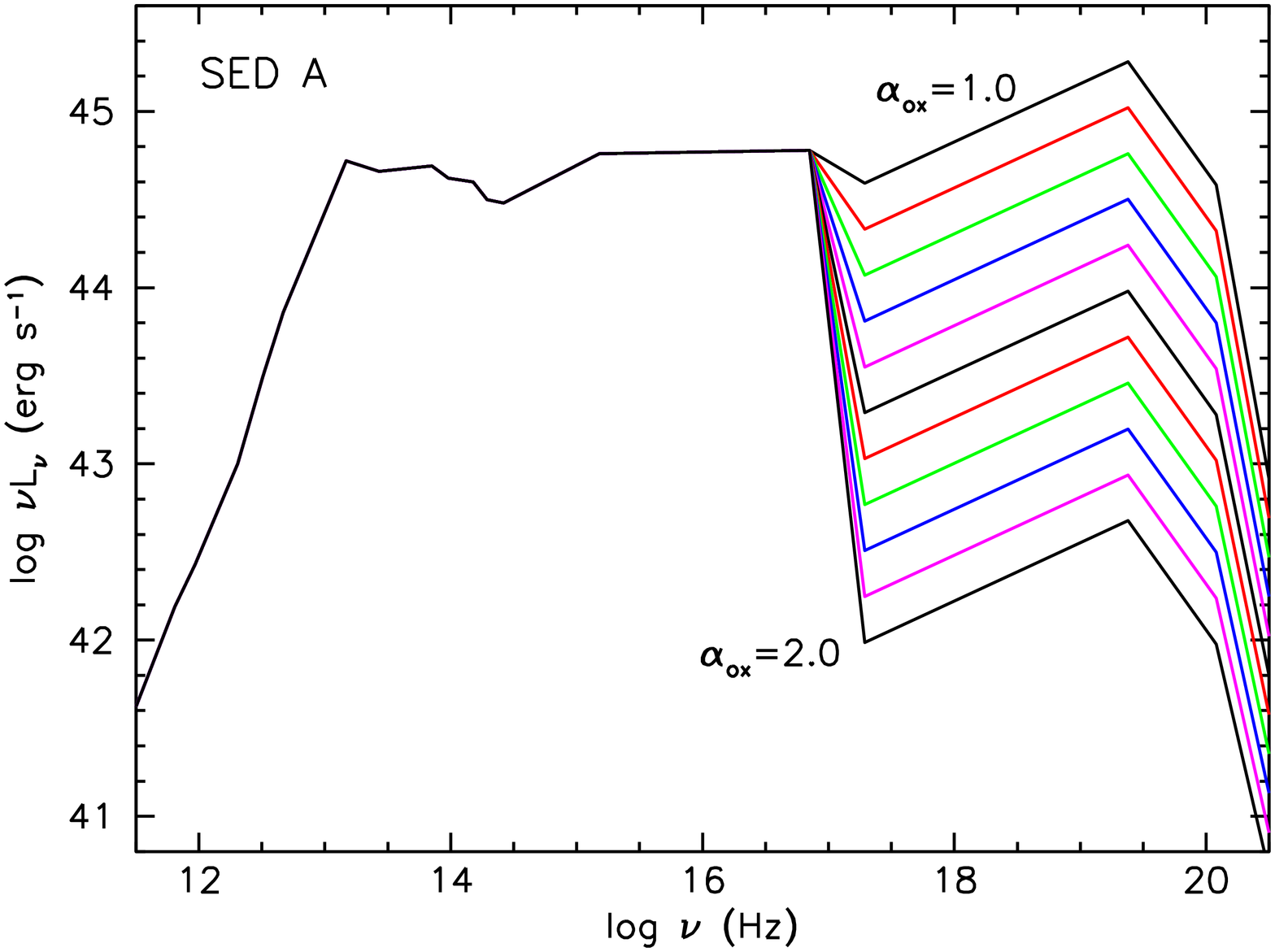}
\includegraphics[angle=-90, width=0.52\textwidth]{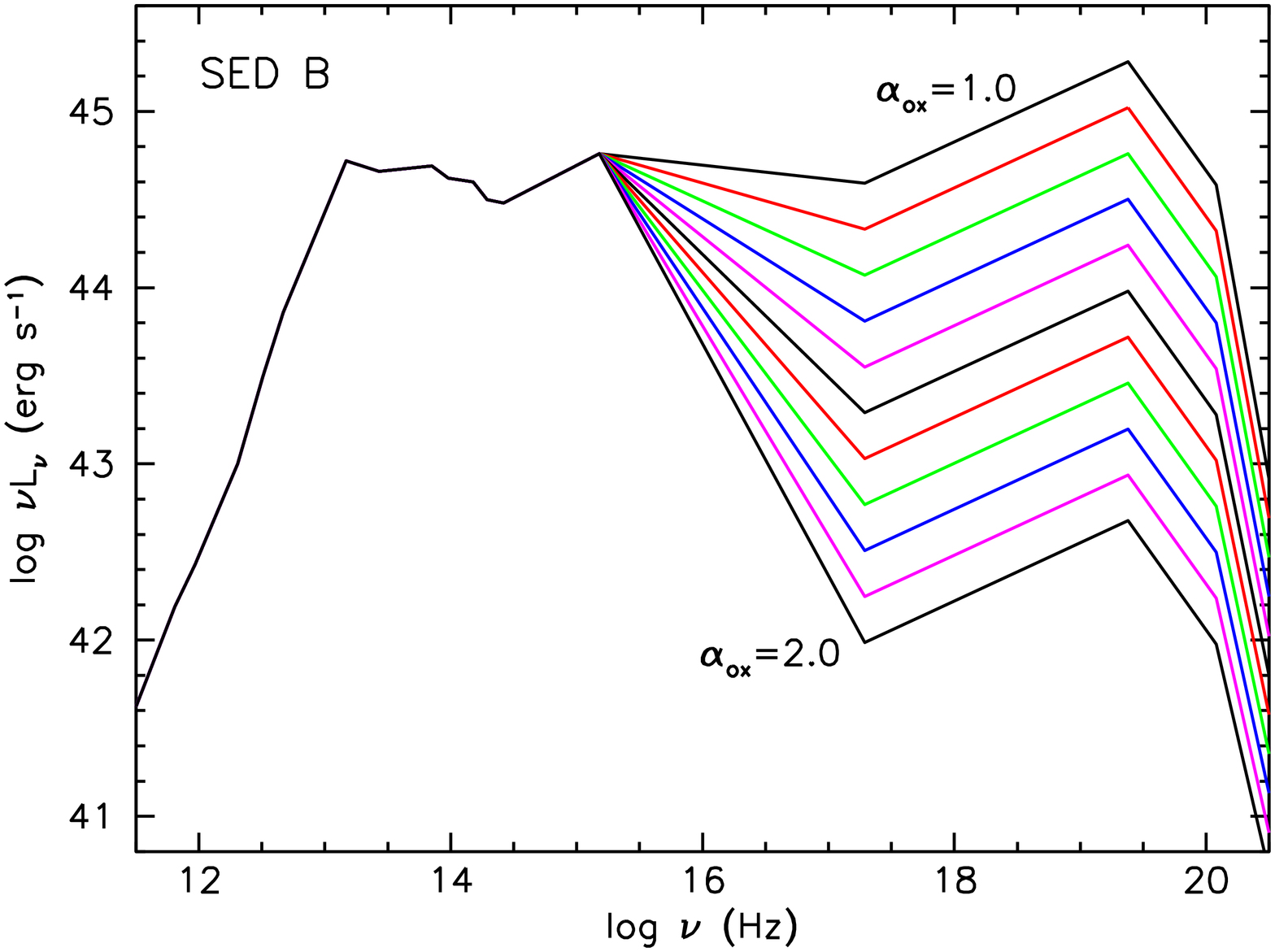}
  \setcounter{figure}{1} 
\caption{
Set of spectral energy distributions used in our photoionization models.
SED A has a strong UV ionizing continuum peaking in the extreme ultraviolet.
Model SEDs in this shape range in $\alpha_{ox}$ from 1.0 to 2.0.
The model with $\alpha_{ox} = 1.5$ is most similar to the SED of
PG 1211+143 used in \cite{Danehkar18}.
SED B has a weaker UV ionizing continuum, with a break from a hard power law
in the UV to a direct extrapolation to the X-ray power law at 1 keV.
Again, model SEDs in this shape range in $\alpha_{ox}$ from 1.0 to 2.0.
}
\label{FigSED}
\end{figure}
\vskip 12pt

\section{Photoionization Models}

The high ionization states implied for UFOs identified with \ion{Fe}{25} or
\ion{Fe}{26} absorption would at first glance suggest that lower-ionization
features such as \ion{H}{1} Ly$\alpha$, \ion{C}{4}, or even \ion{O}{6}
would not be detectable.
However, given the high column densities required to produce detectable
\ion{Fe}{26} Ly$\alpha$ absorption (total $\rm N_H > 10^{23}~\rm cm^{-2}$)
\citep{Tombesi11}, an ionization fraction for neutral hydrogen of $10^{-10}$
would still imply a high enough column density ($>10^{13}~\rm cm^{-2}$) of
neutral hydrogen to produce a detectable UV absorption line.
To quantify the conditions under which one might see Ly$\alpha$ absorption
or other UV absorption lines concurrently with \ion{Fe}{26} absorption,
we computed a set of photoionization models covering a broad range in
spectral energy distribution (SED).
Our calculations were performed with version 17.00 of {\sc Cloudy},
last described by \cite{Ferland17}.

\bigskip
\bigskip
\subsection{Spectral Energy Distributions}

While both \cite{Tombesi11} and \cite{Gofford13} generated detailed
grids of photoionization models to evaluate their X-ray spectra,
their spectral energy distributions were more tailored to their
X-ray spectra. The spectral shape in the ionizing ultraviolet is
of little consequence for the ionization balance of
\ion{Fe}{25} or \ion{Fe}{26}. However, for the purposes of our
study, the relative strength and shape of the  ionizing ultraviolet
spectral range has a large impact on the relative strengths of
any UV spectral features that might be associated with the
X-ray outflow.
(See the analysis of IRAS13349+2438 by \cite{Lee13} for a detailed example.)
For a baseline spectral energy distribution, we started with the observed SED
of PG 1211+143 \citep{Danehkar18}.
As shown in Figure 4 of \cite{Danehkar18}, this SED has a hard ionizing
continuum in the extreme ultraviolet, a strong soft X-ray excess, and a typical
hard X-ray power law.
The UV and extreme UV is characterized by a powerlaw
$f_\nu \sim \nu^{- \alpha}$ with index $\alpha = 0.99$,
which breaks at 0.25 keV
to an index of 5.57 to form the soft X-ray excess.
This soft X-ray power law breaks at 0.8 keV to a hard X-ray power law with
an energy index of 0.67, which is cut off at
100 keV with a break to an energy index of 2.
This spectral shape is similar to the median spectral energy distribution
of quasars, e.g., \cite{Elvis94} or \cite{Shang2011}, although the X-ray
spectral energy index of 0.67 is slightly harder than the mean of the
\cite{Tombesi10} UFO sample, which \cite{Tombesi11} give as 0.8.

Starting with this basic shape, and keeping the hard X-ray (E $>$ 0.8 keV)
and UV portions of the SED the same (i.e., $\alpha = 0.99$ in the UV,
and $\alpha = 0.67$ in the hard X-ray), we adjust the normalization at 0.8 keV
to generate a set of SEDs spanning a range in $\alpha_{ox}$ from
1.0 to 2.0. ($\alpha_{ox}$ is the effective spectral index from 2500 \AA\ in the
UV to 2 keV in the X-ray.)
These SEDs are illustrated in the top panel of Figure \ref{FigSED}, and they
are called ``SED A".

As an alternative, we also explored a set of SEDs that have a much softer
spectrum in the hydrogen-ionizing portion of the spectrum.
These spectra have a break at the Lyman limit (13.6 eV/912 \AA) that
extrapolates directly to the soft X-ray at 0.8 keV.
Such spectra are typical of many AGN, as illustrated by the
$\sim1000$ \AA\ breaks seen in composite UV spectra \citep{Zheng97, Telfer02}.
and in the spectra of many individual quasars \citep{Shang05}.
These soft spectra are also expected to have higher \ion{H}{1} ionization
fractions for a comparable level of ionization fractions in X-ray ionic
species \citep{Kriss18}.
Again, as for SED A, we generate a set of spectra ranging from 1.0 to 2.0
in $\alpha_{ox}$ by adjusting the normalization of the hard X-ray spectrum
at 0.8 keV.
The spectra in the bottom panel of Figure \ref{FigSED} show the spectra
representing ``SED B".

\bigskip
\bigskip
\bigskip
\subsection{X-ray to Optical Luminosity Ratios of the UFO Sample}

Our grid of spectral energy distributions allows us to investigate
the potential properties of \ion{H}{1} Ly$\alpha$ absorption associated
with a wide range of spectra. The objects in our particular sample,
however, are quite well studied, and so we can examine predictions for
each with better precision if we use our knowledge of their
individual SEDs. Although we do not produce specific SEDs for each
individual object (which would still have the uncertainty of
the shape of the unobservable ionizing ultraviolet continuum),
we used literature sources and the NASA Extragalactic Database (NED)
to obtain X-ray to optical luminosity ratios for each object.
We characterize these ratios by the usual effective spectral index
from 2500 \AA\ in the UV to 2 keV in the X-ray, or $\alpha_{ox}$.
Our preference for selection of values were (1) a fully modeled
SED based on simultaneous X-ray and UV observations;
(2) values based on simultaneously measured UV and X-ray fluxes;
(3) values based on a published SED (but not necessarily modeled
using simultaneous observations);
(4) values based on median values from NED;
(5) a calculation using median values from NED, but inferring a
correction for internal absorption in the host galaxy by
normalizing the J-band flux from NED to the median radio-quiet
SED of \cite{Shang2011}.
Table \ref{tab:alpha_ox} gives our adopted $\alpha_{ox}$ for
each source, the method used (1--5 as described above), and
the reference for our data.
Note that methods 3--5 can lead to uncertainties of
$\sim 0.12$ in $\alpha_{ox}$ for relative variations in
flux between the UV and the X-ray of a factor of 2.

\begin{tablehere}
\begin{center}
  \caption[]{X-ray to Optical Luminosity Ratios\\
($\alpha_{ox}$) for the UFO Sample}
  \label{tab:alpha_ox}
\begin{tabular}{l c c c}
\hline\hline
{Object} & $\alpha_{ox}$ & Method & Reference \\
         &               &        &           \\
\hline
1H0419$-$577 & 1.24 & 5 & 1 \\
Ark 120  & 1.32 & 2 & 2 \\
Mrk 79   & 1.40 & 5 & 1 \\
NGC 4051 & 1.48 & 3 & 3 \\
NGC 4151 & 1.07 & 1 & 4 \\
PG 1211+143 & 1.47 & 1 & 5 \\
Mrk 766  & 1.38 & 5 & 1 \\
Mrk 205  & 1.13 & 4 & 1 \\
Mrk 279  & 1.19 & 2 & 2 \\
Mrk 841  & 1.36 & 5 & 1 \\
Mrk 290  & 1.31 & 5 & 1 \\
Mrk 509  & 1.31 & 1 & 6 \\
MR2251$-$178 & 1.14 & 2 & 2 \\
\hline
\end{tabular}
\end{center}
References: (1) NED, (2) \cite{Laha14}, (3) \cite{Kraemer12},
(4) \cite{Kraemer05}, (5) \cite{Danehkar18},
(6) \cite{Mehdipour11}.

\end{tablehere}

\subsection{Photoionization Model Results}

Given the SEDs described above, we ran {\sc Cloudy} models covering the full
range in $\alpha_{ox}$ for each SED.
Each model is characterized by the total hydrogen particle density, $n_H$
(which is fixed at $10^{10}~\rm cm^{-3}$), the total hydrogen column density
$\rm N_H$, and the ionization parameter $\xi$ = $L_{ion} / (n_H r^2)$,
where $L_{ion}$ is the ionizing luminosity
integrated from the Lyman limit to 1000 Rydbergs,
and $r$ is the distance from the central source.
We used the default {\sc Cloudy} solar abundances.
As shown in \cite{Danehkar18}, the relevant photoionization solutions are
insensitive to densities over the likely density range
of $10^8 - 10^{14}~\rm cm^{-3}$.
Each of these models covered a range in ionization parameter $\xi$
from log $\xi = 2.0$ to 5.0 in steps of 0.05 dex.
(We ignored lower ranges of $\xi$ since significant populations of
\ion{Fe}{25} or \ion{Fe}{26}, the basis of our study, are not formed at
lower values of the ionization parameter.)
Given that ionization fronts develop at high column densities and/or low
ionization parameters, we found that it is essential to sample the full range
of total column density.
Therefore, our models also step through a range in the stopping column density
from log $\rm N_H$ = 22.0--24.3 $\rm cm^{-2}$ at 0.1 dex intervals.

\begin{figurehere}
\centering
\includegraphics[width=0.52\textwidth]{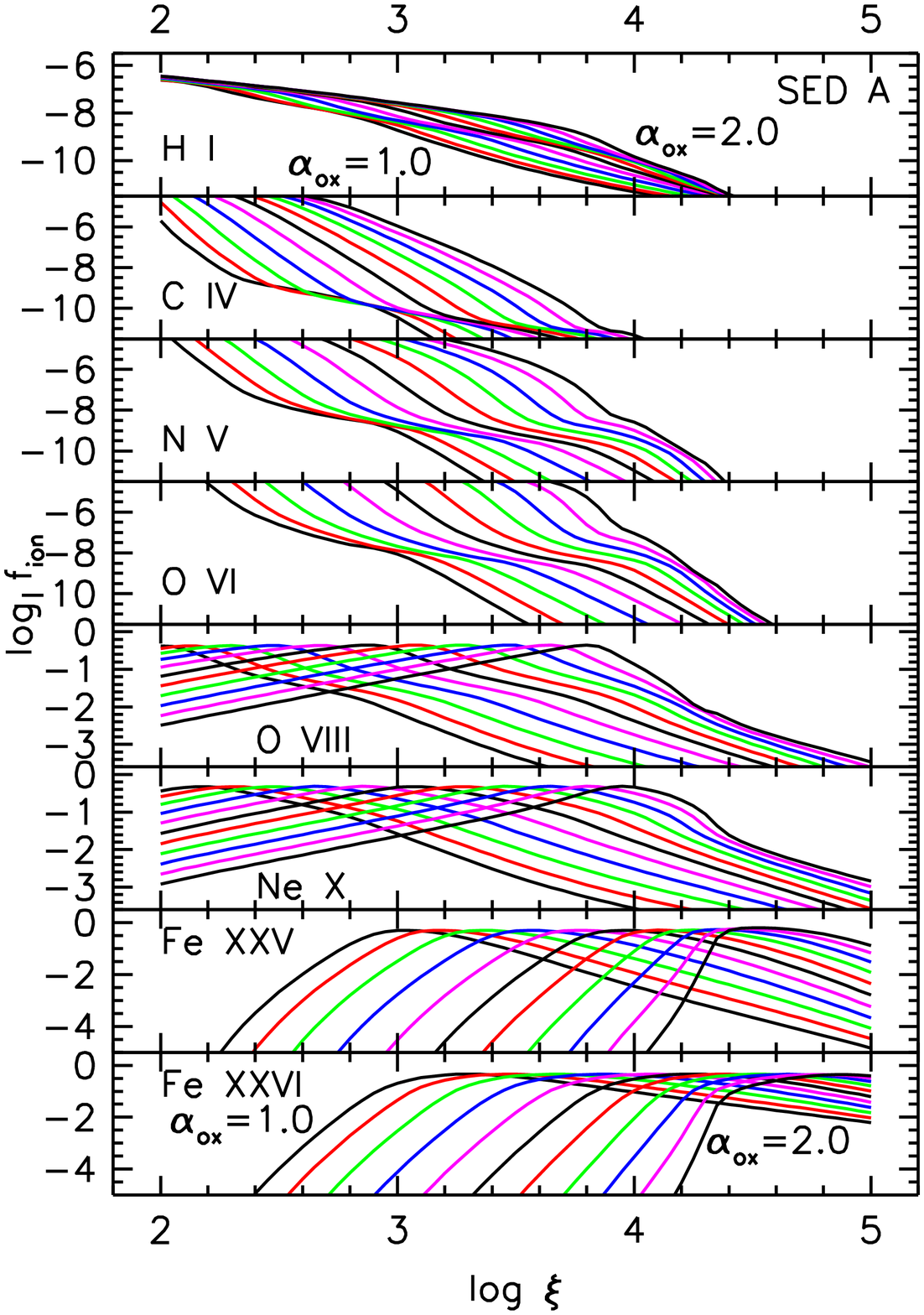}
\caption{
Ionization fractions as a function of ionization parameter for SED A,
which has a hard extreme ultraviolet ionizing flux,
assuming a total column density of log $\rm N_H$ = 23.0 $\rm cm^{-2}$.
The alternating color curves span a range in $\alpha_{ox}$ from 1.0 to
2.0 at intervals of 0.1.
}
\label{FigIonFracsA}
\end{figurehere}

\begin{figurehere}
\centering
\includegraphics[width=0.52\textwidth]{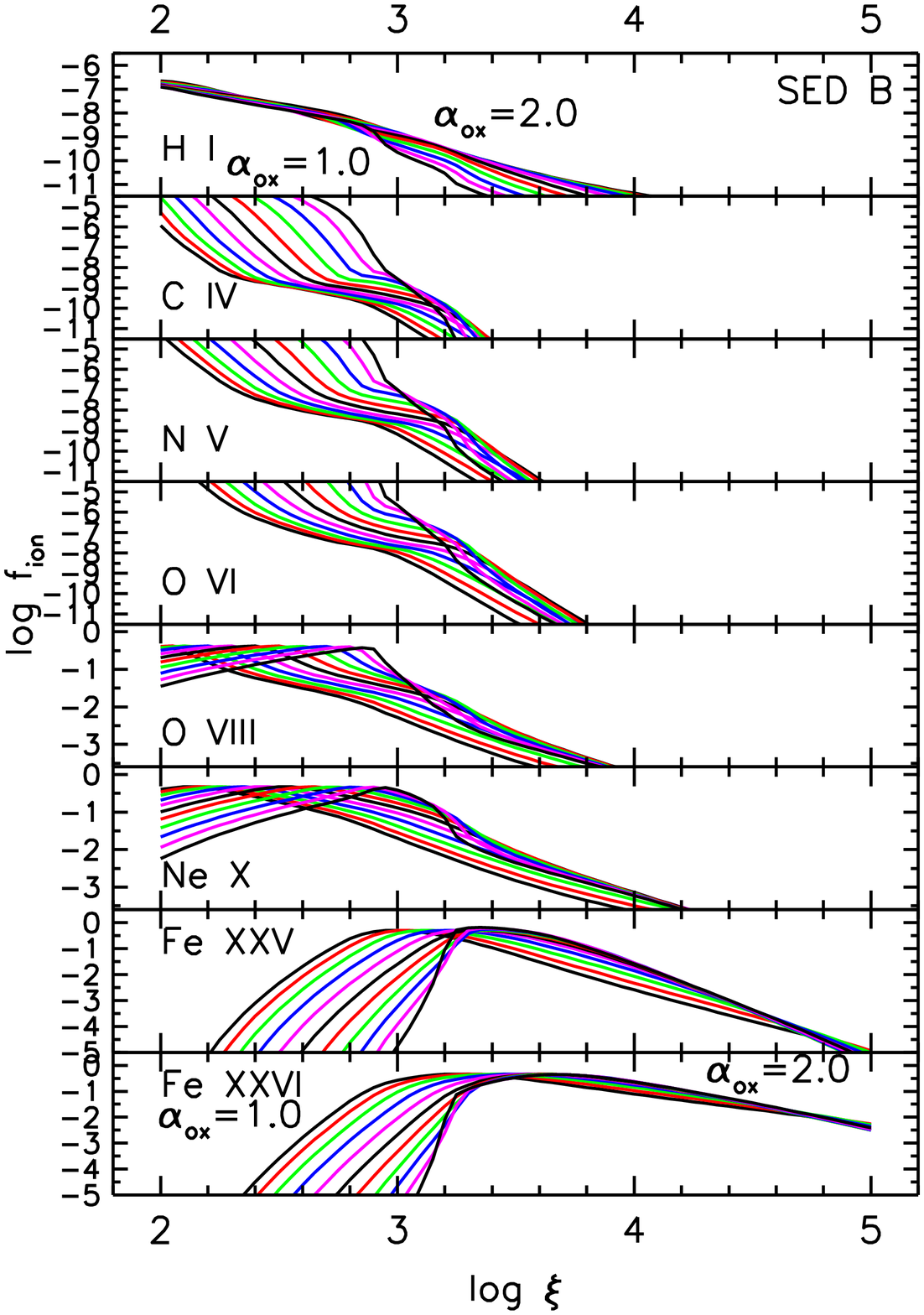}
\caption{
Ionization fractions as a function of ionization parameter for SED B,
which has a soft extreme ultraviolet ionizing flux,
assuming a total column density of log $\rm N_H$ = 23.0 $\rm cm^{-2}$.
The alternating color curves span a range in $\alpha_{ox}$ from 1.0 to
2.0 at intervals of 0.1.
}
\label{FigIonFracsB}
\end{figurehere}
\vskip 12pt

Figures \ref{FigIonFracsA} and \ref{FigIonFracsB} show
ionization fractions as a function of log $\xi$
for the main transitions of interest for our study,
\ion{H}{1}, \ion{C}{4}, \ion{N}{5}, \ion{O}{6},
\ion{Fe}{25}, and \ion{Fe}{26},
computed for a fixed column density of log $\rm N_H$ = 23.0 $\rm cm^{-2}$,
which is typical of the gas forming \ion{Fe}{26} Ly$\alpha$ absorption
features \citep{Tombesi11}.
Note the inflections in the fractional ionization curves for the
lower-ionization ions that form ``ripples" across the figures.
These are due to the ionization fronts that develop for higher-ionization ions
at ionization parameters that make the calculation ionization bounded for
a total column density of log $\rm N_H$ = 23.0 $\rm cm^{-2}$ for those ions.

We also show representative intermediate ionization states, namely
\ion{O}{8} and \ion{Ne}{10}, since lower ionization parameters that favor
higher observable columns of \ion{H}{1} might also be expected to produce
lower-energy X-ray spectral lines easily observed with grating instruments.
As anticipated, peak ionization fractions for \ion{Fe}{25} and \ion{Fe}{26}
require ionization parameters between log $\xi$ = 3.0 and 4.5,
and corresponding \ion{H}{1} ionization fractions are quite low, in the range
of $10^{-8}$ to $10^{-10}$.
However, as noted earlier, the dominant overall abundance of hydrogen
can still produce detectable UV absorption lines with column densities
of order $10^{13}~\rm cm^{-2}$.
\ion{C}{4}, \ion{N}{5}, and \ion{O}{6} have higher ionization fractions than
hydrogen, but, given that their abundances are lower by factors of several
thousand, they ultimately have lower predicted column densities than
\ion{H}{1} itself.

\begin{figurehere}
\centering
\includegraphics[width=0.52\textwidth]{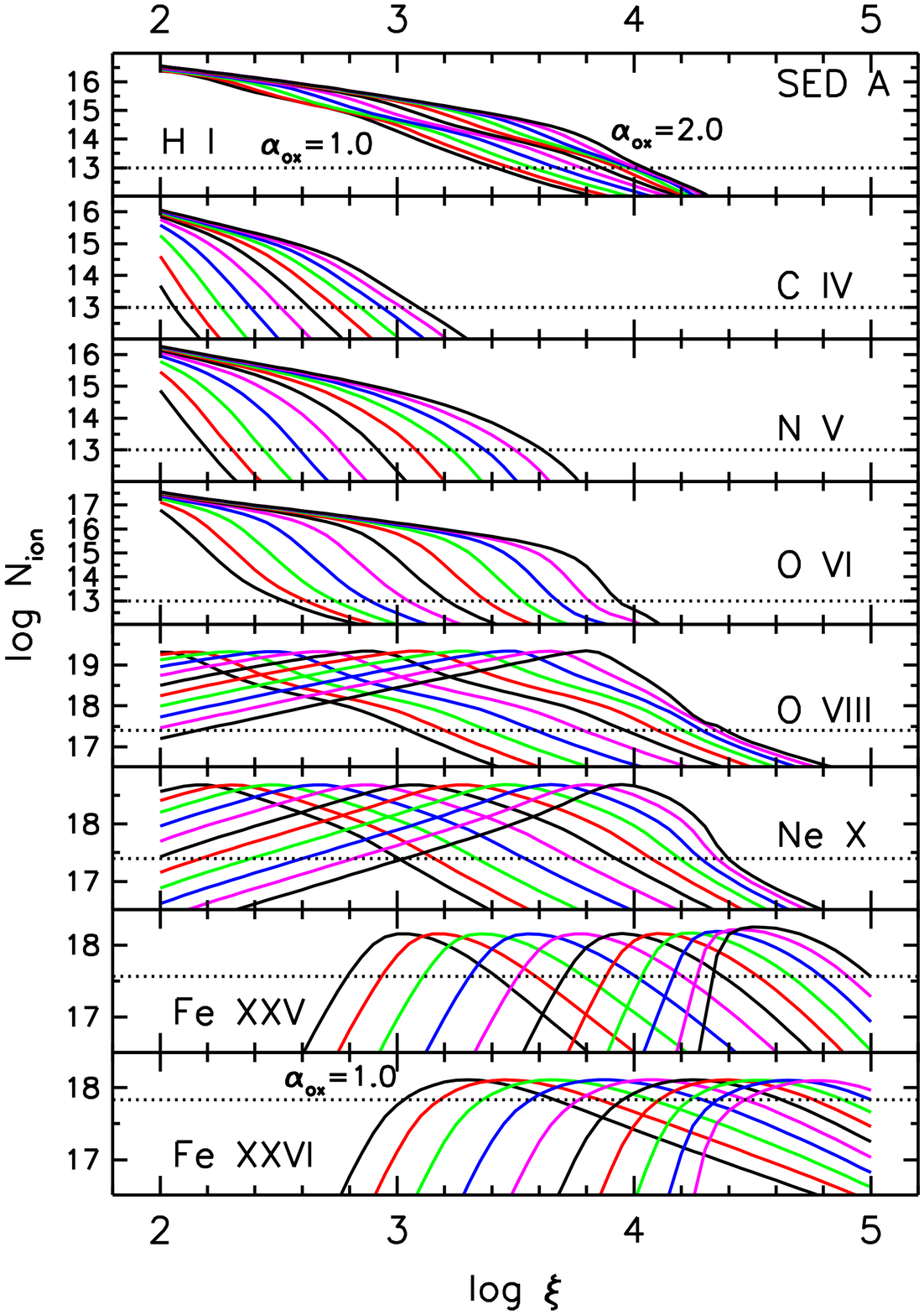}
\caption{
Ionic column densities as a function of ionization parameter for SED A,
which has a hard extreme ultraviolet ionizing flux,
assuming a total column density of log $\rm N_H$ = 23.0 $\rm cm^{-2}$.
The alternating color curves span a range in $\alpha_{ox}$ from 1.0 to
2.0 at intervals of 0.1.
Dashed horizontal lines show approximate thresholds for each ion for detecting
UV or X-ray absorption lines (see text).
}
\label{FigNionA}
\end{figurehere}

\begin{figurehere}
\centering
\includegraphics[width=0.52\textwidth]{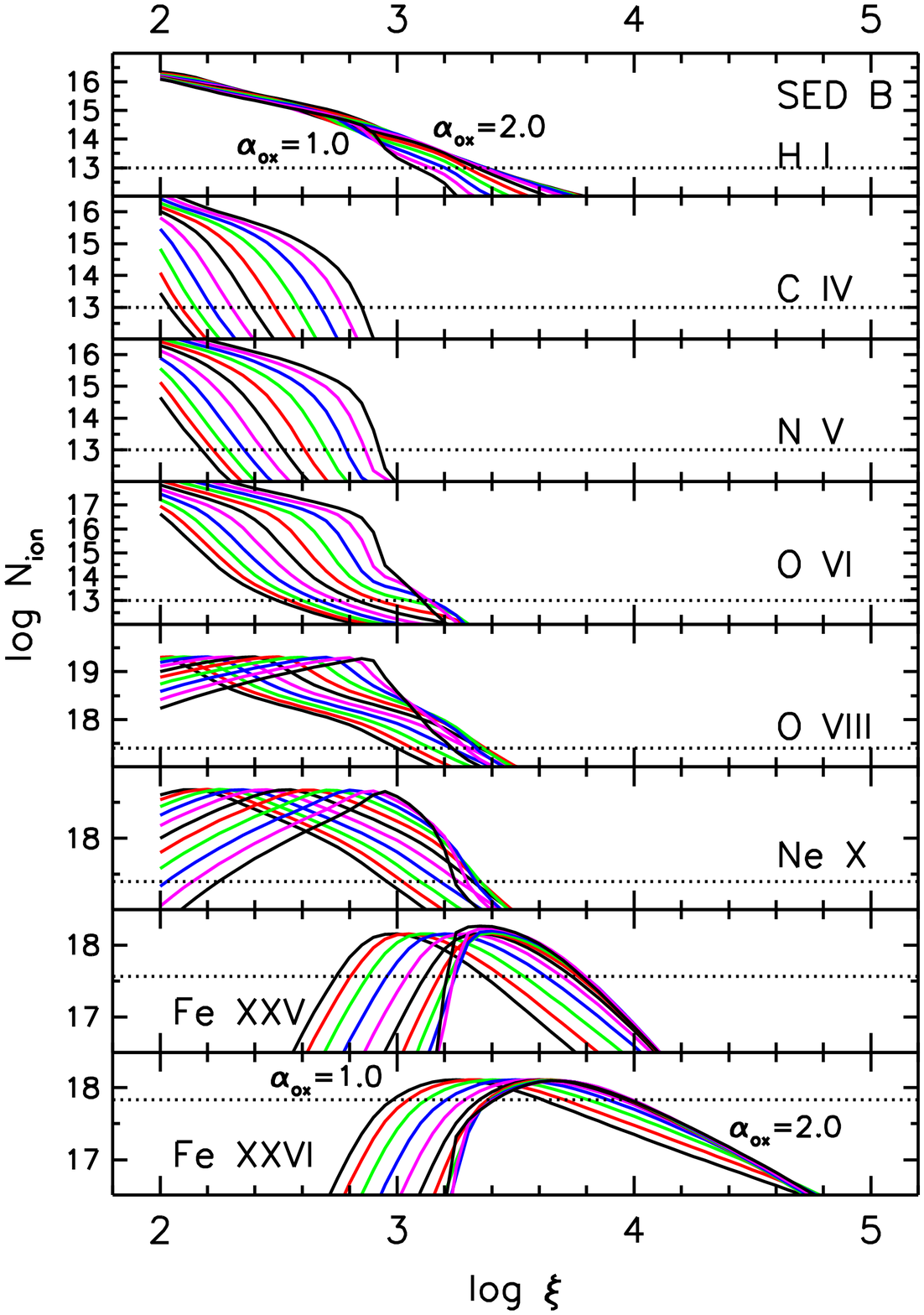}
\caption{
Ionic column densities as a function of ionization parameter for SED B,
which has a soft extreme ultraviolet ionizing flux,
assuming a total column density of log $\rm N_H$ = 23.0 $\rm cm^{-2}$.
The alternating color curves span a range in $\alpha_{ox}$ from 1.0 to
2.0 at intervals of 0.1.
Dashed horizontal lines show approximate thresholds for each ion for detecting
UV or X-ray absorption lines (see text).
}
\label{FigNionB}
\end{figurehere}
\vskip 12pt

Figures \ref{FigNionA} and \ref{FigNionB} show the individual ionic column
densities as a function of ionization parameter corresponding to the ionic
fractions for a total column density of
log $\rm N_H$ = 23.0 $\rm cm^{-2}$.
To interpret these curves in terms of observable quantities, we must assume
a velocity width for the spectral features of interest.
Given the $1000~\rm km~s^{-1}$ width of the Ly$\alpha$ feature in
PG 1211+143 \citep{Kriss18}, at $\rm log~N_{ion} < 13.0~cm^{-2}$ all UV features
are optically thin. \ion{H}{1}, \ion{C}{4}, \ion{N}{5}
and \ion{O}{6} are typically undetectable if log $\rm N_{ion} < 13.0 ~cm^{-2}$,
and these thresholds are indicated in the figures.
This breadth must be interpreted as a turbulent velocity.
In our photoionization models, temperatures are roughly $10^6$ K,
with a thermal width for \ion{H}{1} of $120~\rm km~s^{-1}$,
and $17~\rm km~s^{-1}$ for Fe.
The UFO features in the CCD spectra of \cite{Tombesi10} and \cite{Gofford13}
are unresolved, and could easily have velocity widths of several thousand
$\rm km~s^{-1}$. The better studied UFOs in PDS 456 \citep{Reeves18} and
APM 08279+5255 \citep{Saez11} are resolved at widths of tens of thousands of
$\rm km~s^{-1}$.
The \ion{C}{4} broad-absorption-line troughs in APM 08279+5255 also have
widths of 2000--2500 $\rm km~s^{-1}$.
As one can see in the curves of growth computed by \citep{Tombesi11}, at
line widths above $\sim 1000~\rm km~s^{-1}$, the Fe features are close to
optically thin.
We will therefore assume they are optically thin for our analysis.
At peak ionic fraction for \ion{Fe}{26}, given a total column density of
$10^{23}~\rm cm^{-2}$, the predicted EW for an \ion{Fe}{26} Ly$\alpha$
absorption feature is 55 eV, which is a common strength for many UFOs.
Limiting equivalent widths of 30-eV EW in \ion{Fe}{25}
and \ion{Fe}{26} require log $\rm N_{ion} = 17.57$ and 17.84, respectively.
For \ion{O}{8} and \ion{Ne}{10}, grating observations can usually detect lines
with column densities log $\rm N_{ion} > 17.4~cm^{-2}$.
Thus, for the soft SED B, if the X-ray UFO absorption feature is
\ion{Fe}{25} or \ion{Fe}{26}, one generally would not expect to see any
lower ionization features in either the UV or the X-ray except for
perhaps Ly$\alpha$.
For the harder SED A, however, one might see lower-ionization X-ray ions such
as \ion{O}{8} and \ion{Ne}{10} if $\alpha_{ox} < 1.5$, but again,
no UV ions other than Ly$\alpha$ would typically be visible.

\begin{figure}[t]
\centering
\includegraphics[angle=-90, width=0.52\textwidth]{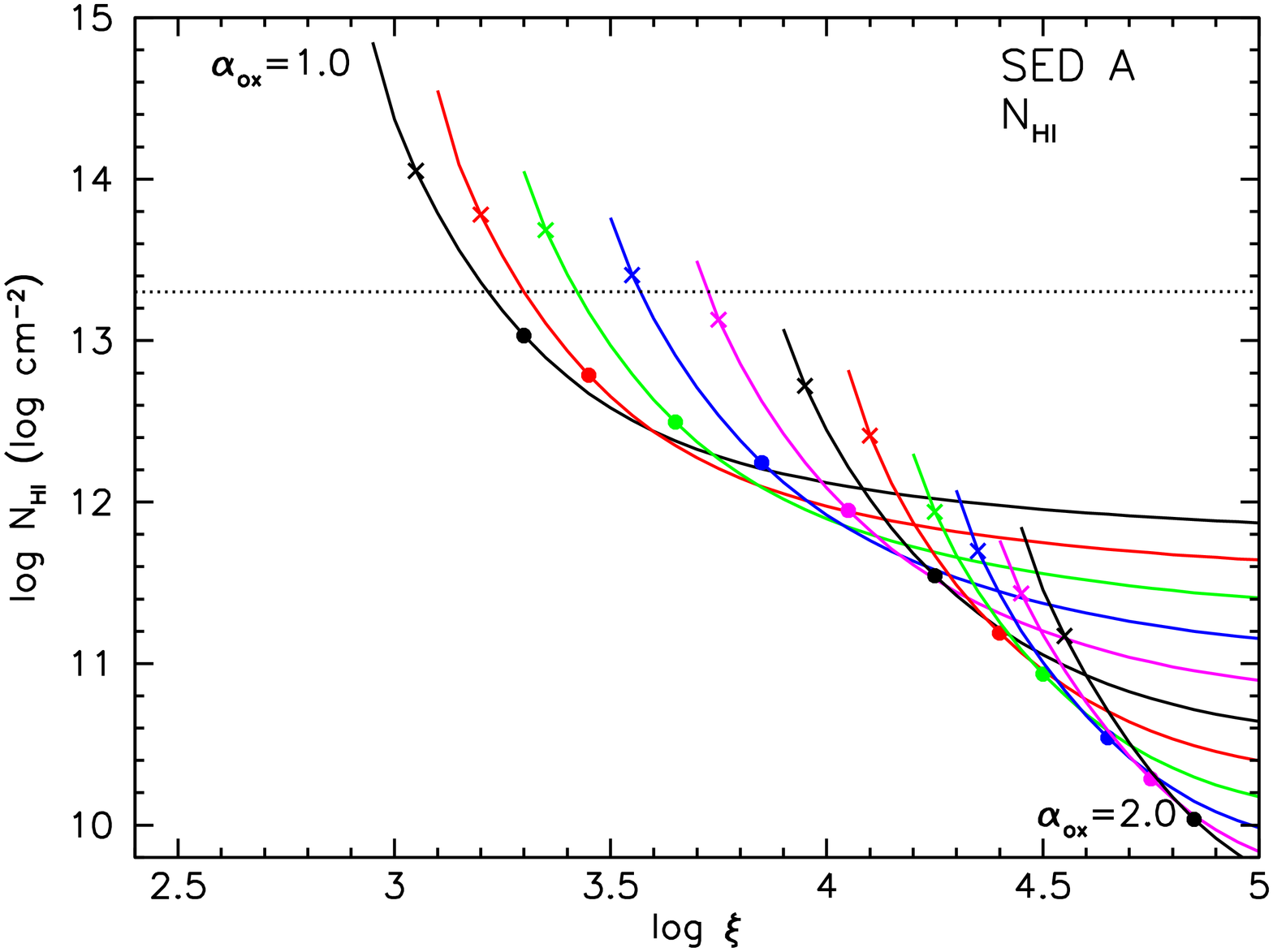}
\includegraphics[angle=-90, width=0.52\textwidth]{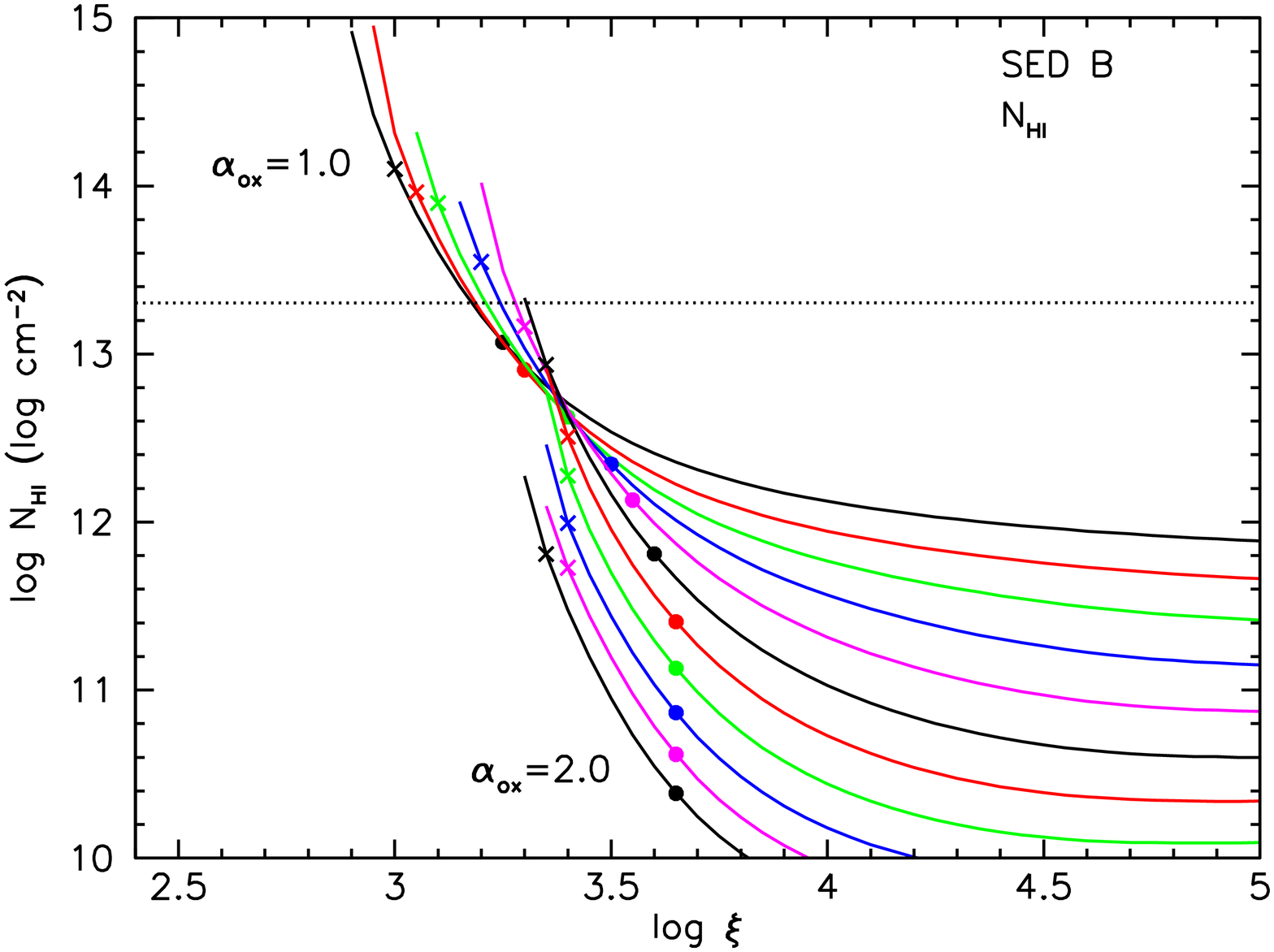}
\caption{
Predicted neutral hydrogen column densities for photoionization
models producing \ion{Fe}{26} absorption with EW=30 eV.
The models in Panel A use SED A with $\alpha_{ox}$ ranging from 1.0 to 2.0.
The models in Panel B use SED B with $\alpha_{ox}$ ranging from 1.0 to 2.0.
Solid dots along the curves mark the ionization parameters for which the
ionization fraction of \ion{Fe}{26} is at a maximum.
Crosses give the peak ionization fraction for \ion{Fe}{25}.
}
\label{FigPredictedHIFe26}
\end{figure}

\begin{figure}[t]
\centering
\includegraphics[angle=-90, width=0.52\textwidth]{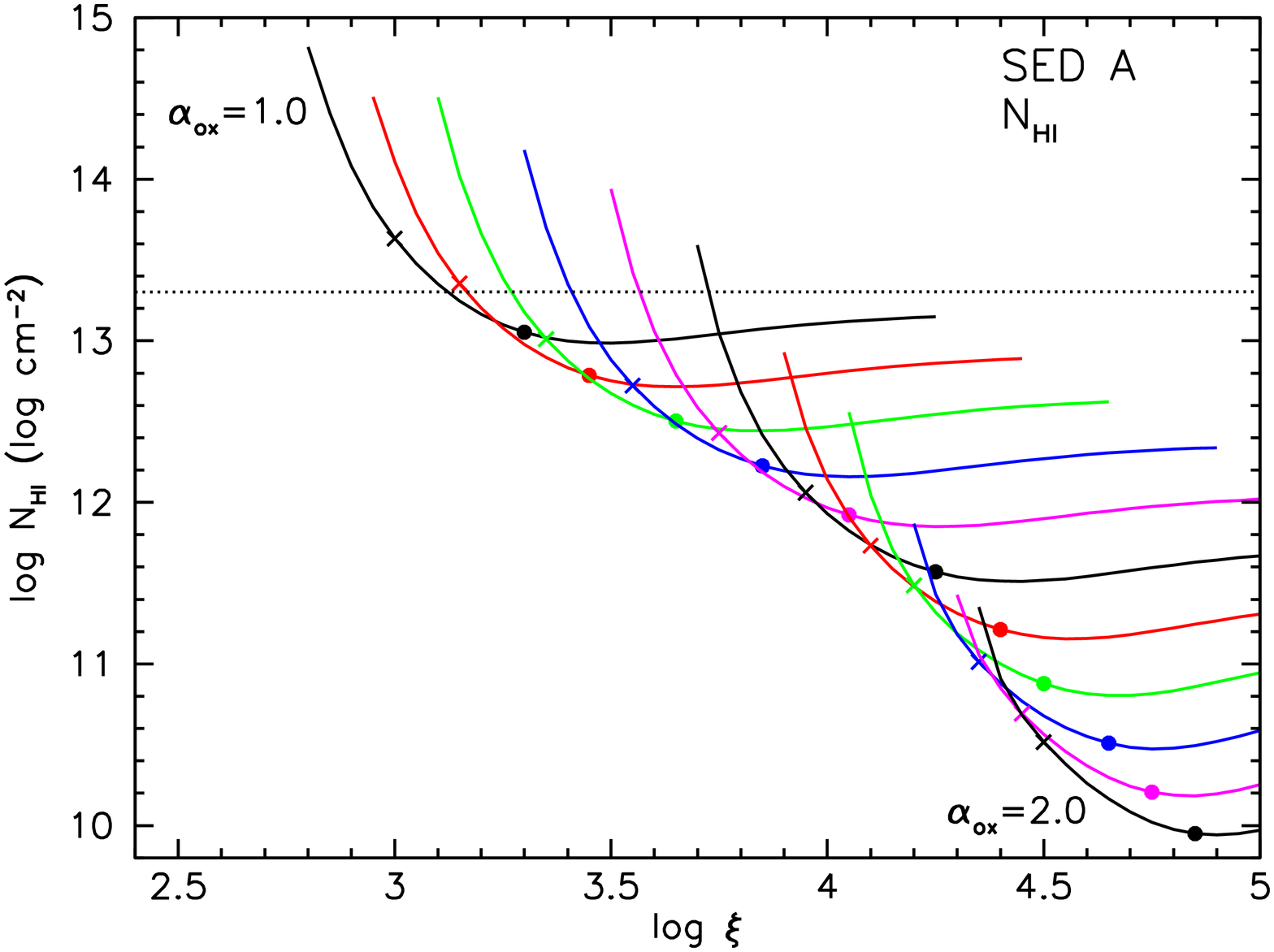}
\includegraphics[angle=-90, width=0.52\textwidth]{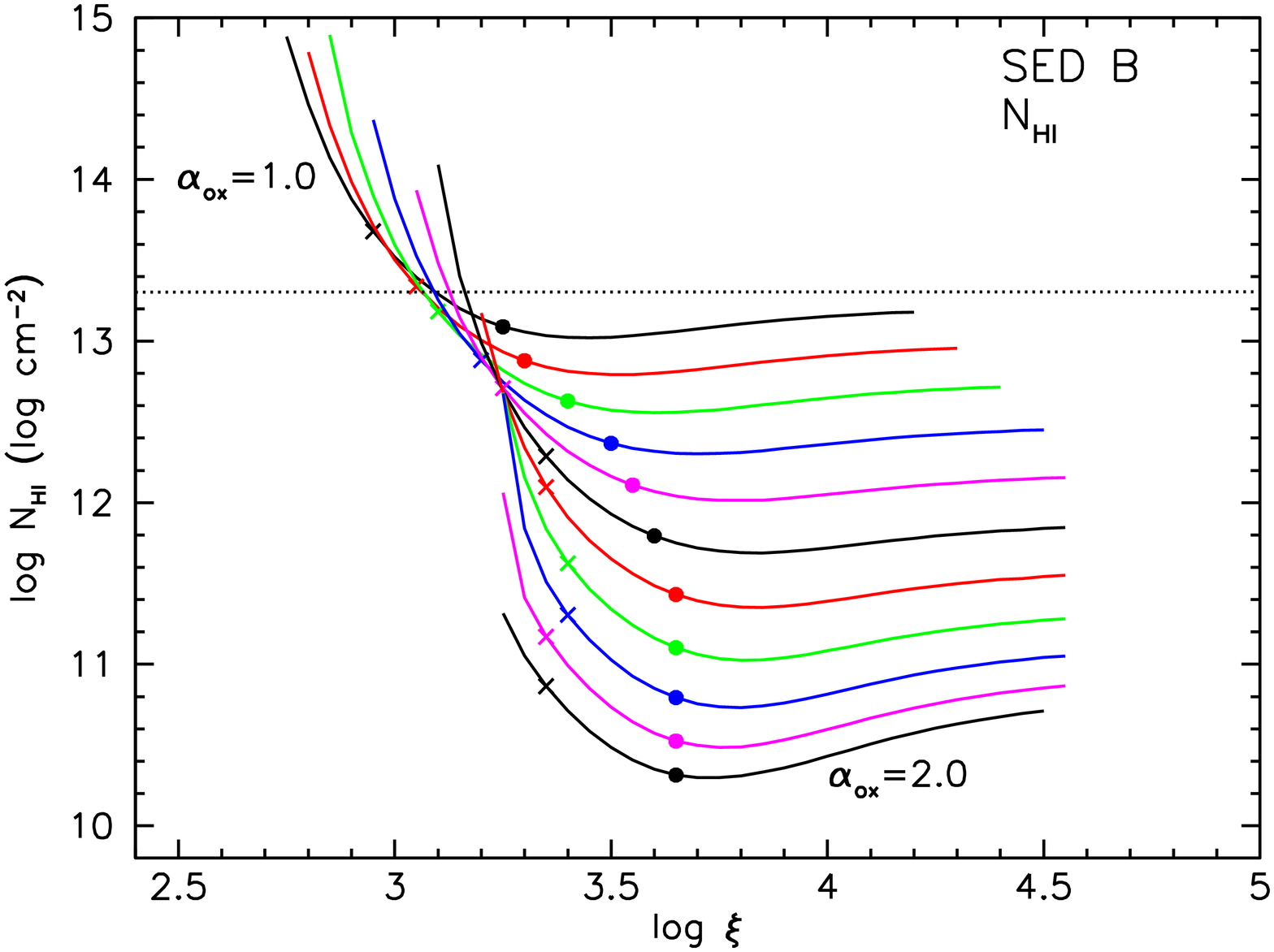}
\caption{
Predicted neutral hydrogen column densities for photoionization
models producing \ion{Fe}{25} absorption with EW=30 eV.
The models in Panel A use SED A with $\alpha_{ox}$ ranging from 1.0 to 2.0.
The models in Panel B use SED B with $\alpha_{ox}$ ranging from 1.0 to 2.0.
Solid dots along the curves mark the ionization parameters for which the
ionization fraction of \ion{Fe}{25} is at a maximum.
Crosses give the peak ionization fraction for \ion{Fe}{25}.
}
\label{FigPredictedHIFe25}
\end{figure}

Using our grid of photoionization models, we can evaluate the detectability
of UV absorption features associated with our sample of UFO outflows.
Tables \ref{tab:Predictions26} and \ref{tab:Predictions25}
give predicted \ion{H}{1} column densities for
each source, given its $\alpha_{ox}$ and the EW of its Fe absorption feature.
In most cases, the predicted \ion{H}{1} column densities are well below the
upper limits given in Table \ref{tab:limits}.
However, as one can see in Figures \ref{FigIonFracsA} and \ref{FigIonFracsB},
\ion{Fe}{25} and \ion{Fe}{26} have high ionization fractions extending
to much lower ionization parameters than at peak.
To ascertain more generally what spectral shapes and ionization parameters
would produce both detectable Fe absorption lines and UV absorption lines,
we computed the resulting ionic column densities set by the requirements that
the EW of \ion{Fe}{25} or \ion{Fe}{26} K$\alpha$ transitions produce an
X-ray absorption feature of 30 eV
(a typical detection level in \cite{Tombesi10}).
Given these requirements on the Fe column densities, for each ionization
parameter in our model grid we identified the corresponding column densities
of all other ions in the model grid.
The predictions for \ion{H}{1} given a 30 eV \ion{Fe}{26} feature
are shown in Figure \ref{FigPredictedHIFe26}, and in
Figure \ref{FigPredictedHIFe25} for \ion{Fe}{25}.
The solid dots in the figure show the column densities predicted at peak ion
fractions for \ion{Fe}{26}, and crosses give the peak ionization fractions
for \ion{Fe}{25}. These are generally lower than an ``easy to detect" limit
of $\rm N_{HI} = 2 \times 10^{13}~\rm cm^{-2}$ for a Ly$\alpha$ feature.
However, for SEDs at low $\alpha_{ox}$ (i.e., $\sim$1.2) and lower-than-peak
ionization fractions for Fe, Ly$\alpha$ features are certainly detectable.
All other UV ions, however, remain at predicted column densities below
$\rm N_{ion} = 10^{12}~\rm cm^{-2}$, which are undetectable;
therefore we do not show any curves for these ions.

\vskip 12pt

\vskip 12pt

\begin{table*}
  \caption[]{Predicted \ion{H}{1} Column Densities for X-ray UFOs in \ion{Fe}{26}}
  \label{tab:Predictions26}
\begin{center}
\begin{tabular}{l l c c c c c c}
\hline\hline
{Object} & EW & log $\rm N_{H}$ & $\alpha_{ox}$ & $\xi_{peak-A}$ & $\xi_{peak-B}$ & $\rm log~N_{HI-A}$ & $\rm log~N_{HI-B}$\\
      &   (eV)    & ($\rm log~cm^{-2}$) &     &  ($\rm log~cm^{-2}$)\\
\hline
1H0419$-$577  & 55 & 22.99 & 1.24 & 3.64 & 3.38 & 12.78 & 12.91 \\
Ark 120  & 25 & 22.65 & 1.32 & 3.84 & 3.48 & 12.16 & 12.26 \\
Mrk 79   & 43 & 22.88 & 1.40 & 4.04 & 3.54 & 12.12 & 12.30 \\
NGC 4051 & 96 & 23.23 & 1.48 & 4.25 & 3.65 & 12.11 & 12.23 \\
NGC 4151 & 28 & 22.70 & 1.07 & 3.44 & 3.30 & 12.75 & 12.87 \\
         & 32 & 22.75 & 1.07 & 3.44 & 3.30 & 12.82 & 12.94 \\
PG 1211+143 & 130 & 23.36 & 1.47 & 4.25 & 3.65 & 12.27 & 12.40 \\
Mrk 766  & 39 & 22.84 & 1.38 & 4.04 & 3.54 & 12.07 & 12.26 \\
Mrk 205  & 50 & 22.95 & 1.13 & 3.44 & 3.30 & 13.02 & 13.15 \\
Mrk 279  & 73 & 23.11 & 1.19 & 3.64 & 3.38 & 12.92 & 13.05 \\
Mrk 841  & 46 & 22.91 & 1.36 & 4.04 & 3.54 & 12.15 & 12.33 \\
Mrk 290  & 85 & 23.18 & 1.31 & 3.90 & 3.48 & 12.62 & 12.84 \\
Mrk 509  & 32 & 22.75 & 1.31 & 3.84 & 3.48 & 12.27 & 12.37 \\
         & 19 & 22.53 & 1.31 & 3.84 & 3.48 & 12.04 & 12.14 \\
         & 19 & 22.53 & 1.31 & 3.84 & 3.48 & 12.04 & 12.14 \\
MR2251$-$178& 32 & 22.75 & 1.14 & 3.44 & 3.30 & 12.82 & 12.94 \\
\hline
\end{tabular}
\end{center}
{\bf Notes.} (1) Object name. (2) Equivalent width of the UFO.
(3) Total hydrogen column density required to produce a feature of that EW
at the peak ion fraction.
(4) $\alpha_{ox}$ for the object.
(5) Ionization parameter required to achieve peak ion fraction given
$\alpha_{ox}$ for SED A.
(6) Ionization parameter required to achieve peak ion fraction given
$\alpha_{ox}$ for SED B.
(7) Predicted \ion{H}{1} column density for parameters corresponding to SED A.
(8) Predicted \ion{H}{1} column density for parameters corresponding to SED B.
\end{table*}

\begin{table*}
  \caption[]{Predicted \ion{H}{1} Column Densities for X-ray UFOs in \ion{Fe}{25}}
  \label{tab:Predictions25}
\begin{center}
\begin{tabular}{l l c c c c c c}
\hline\hline
{Object} & EW & log $\rm N_{H}$ & $\alpha_{ox}$ & $\xi_{peak-A}$ & $\xi_{peak-B}$ & $\rm log~N_{HI-A}$ & $\rm log~N_{HI-B}$\\
      &   (eV)    & ($\rm log~cm^{-2}$) &     &  ($\rm log~cm^{-2}$)\\
\hline
1H0419$-$577  & 55 & 22.67 & 1.24 & 3.34 & 3.10 & 13.29 & 13.46 \\
Ark 120  & 25 & 22.33 & 1.32 & 3.54 & 3.18 & 12.64 & 12.80 \\
Mrk 79   & 43 & 22.56 & 1.40 & 3.74 & 3.26 & 12.60 & 12.88 \\
NGC 4051 & 96 & 22.91 & 1.48 & 3.95 & 3.35 & 12.61 & 12.84 \\
NGC 4151 & 28 & 22.38 & 1.07 & 3.15 & 3.05 & 13.32 & 13.31 \\
         & 32 & 22.44 & 1.07 & 3.16 & 3.05 & 13.39 & 13.37 \\
PG 1211+143 & 130 & 23.05 & 1.47 & 3.95 & 3.35 & 12.77 & 13.00 \\
Mrk 766  & 39 & 22.52 & 1.38 & 3.74 & 3.26 & 12.55 & 12.83 \\
Mrk 205  & 50 & 22.63 & 1.13 & 3.16 & 3.05 & 13.59 & 13.57 \\
Mrk 279  & 73 & 22.80 & 1.19 & 3.34 & 3.10 & 13.42 & 13.60 \\
Mrk 841  & 46 & 22.59 & 1.36 & 3.74 & 3.26 & 12.63 & 12.91 \\
Mrk 290  & 85 & 22.86 & 1.31 & 3.54 & 3.18 & 13.21 & 13.37 \\
Mrk 509  & 32 & 22.44 & 1.31 & 3.54 & 3.18 & 12.75 & 12.91 \\
         & 19 & 22.21 & 1.31 & 3.54 & 3.18 & 12.52 & 12.67 \\
         & 19 & 22.21 & 1.31 & 3.54 & 3.18 & 12.52 & 12.67 \\
MR2251$-$178 & 32 & 22.44 & 1.14 & 3.16 & 3.15 & 13.39 & 13.39 \\
\hline
\end{tabular}
\end{center}
{\bf Notes.} (1) Object name. (2) Equivalent width of the UFO.
(3) Total hydrogen column density required to produce a feature of that EW
at the peak ion fraction.
(4) $\alpha_{ox}$ for the object.
(5) Ionization parameter required to achieve peak ion fraction given
$\alpha_{ox}$ for SED A.
(6) Ionization parameter required to achieve peak ion fraction given
$\alpha_{ox}$ for SED B.
(7) Predicted \ion{H}{1} column density for parameters corresponding to SED A.
(8) Predicted \ion{H}{1} column density for parameters corresponding to SED B.
\end{table*}

\newpage
\section{Discussion}

The discovery of \ion{H}{1} Ly$\alpha$ absorption associated with the
$v_{out} = -17\,300~\rm km~s^{-1}$ X-ray UFO in the quasar PG 1211+143
offers the prospect of additional spectral diagnostics for studying
these energetic phenomena.
Such an additional spectral line in a completely different waveband
permits a more detailed assessment of the shape of the overall
ionizing spectrum and the ionization parameter of the gas in the outflow.
Our search for Ly$\alpha$ counterparts in a large sample of X-ray UFOs
sets some fairly stringent upper limits on the presence of \ion{H}{1}
in these fast outflows.

From our large grid of photoionization models and our lack of
Ly$\alpha$ absorption detections, we can conclude that either
the UFOs in this study are infrequently present, or that their
ionization parameters are high.
If the X-ray absorption features detected by \cite{Tombesi10}
and \cite{Gofford13} are produced by \ion{Fe}{26}, as shown in
Table \ref{tab:Predictions26} and Figure \ref{FigPredictedHIFe26},
the predicted \ion{H}{1} column
density falls below our upper limits for all objects.
However, if the X-ray absorption is produced by \ion{Fe}{25},
as illustrated by Table \ref{tab:Predictions25} and
Figure \ref{FigPredictedHIFe25},
in many cases we would have expected to see Ly$\alpha$
absorption if \ion{Fe}{25} was near peak ionization.
Thus we conclude that it is more likely that the X-ray
absorption features are produced by \ion{Fe}{26}.
Furthermore, ionization parameters for a wide range of
models then needs to be higher than log $\xi$=3.2, or
Ly$\alpha$ absorption could be produced at detectable levels
at below-peak ionization fractions for \ion{Fe}{26} if
the column density is high enough,
e.g., $> 5 \times 10^{23}~\rm cm^{-2}$.
Such high ionization parameters are generally expected based
on the photoionization models and fits to the X-ray spectra
performed by \cite{Tombesi11} and \cite{Gofford13}, but this study
does provide independent confirmation.

Our analysis above has rested on assuming that the UFO-producing gas has
large turbulent velocities, $\gtsim 1000~\rm km~s^{-1}$, rendering it
optically thin. If the UFO features do have intrinsic widths of only
100-200 $\rm km~s^{-1}$, as in the lower-ionization UFO in
PG 1211+143 at $v_{out} = -17\,300~\rm km~s^{-1}$ \citep{Danehkar18},
then the required total column densities to
produce the Fe absorption would be significantly higher
\citep[see the curves of growth in][]{Tombesi11}, and the corresponding
predictions for \ion{H}{1} would also be higher.
That would then lead to the prediction that Ly$\alpha$ features would be
even easier to detect, making it surprising that we do not find a
significant number.

Another interesting way to explore our photoionization models is to approach
them from the reverse direction, and investigate what range of
physical conditions can produce strong Ly$\alpha$ absorption as seen by
\cite{Kriss18}.
Figures \ref{FigNtotA} and \ref{FigNtotB} show column densities for the
primary UV and X-ray ions we have been investigating as a function of
ionization parameter for photoionization solutions that would produce
a neutral hydrogen column density of log $\rm N_{HI} = 14.29~\rm cm^{-2}$.
The top panel in each figure shows the total column density associated with
those models.
Such a neutral hydrogen column would produce a
Ly$\alpha$ absorption line with FWHM = $1000~\rm km~s^{-1}$, comparable to
the strength of the Ly$\alpha$ absorption in PG 1211+143, and the fiducial
model we show in the spectra in Figure \ref{FigFitsa}.
As in Figures \ref{FigNionA} and \ref{FigNionB},
\ion{C}{4}, \ion{N}{5} and \ion{O}{6} are typically undetectable if
log $\rm N_{ion} < 13.0 ~cm^{-2}$,
a 30-eV EW feature in \ion{Fe}{25}
or \ion{Fe}{26} requires log $\rm N_{ion} = 17.57$ and 17.84, respectively, and
\ion{O}{8} and \ion{Ne}{10} are visible only
with column densities log $\rm N_{ion} > 17.4~cm^{-2}$.

\begin{figurehere}
\centering
\includegraphics[width=0.52\textwidth]{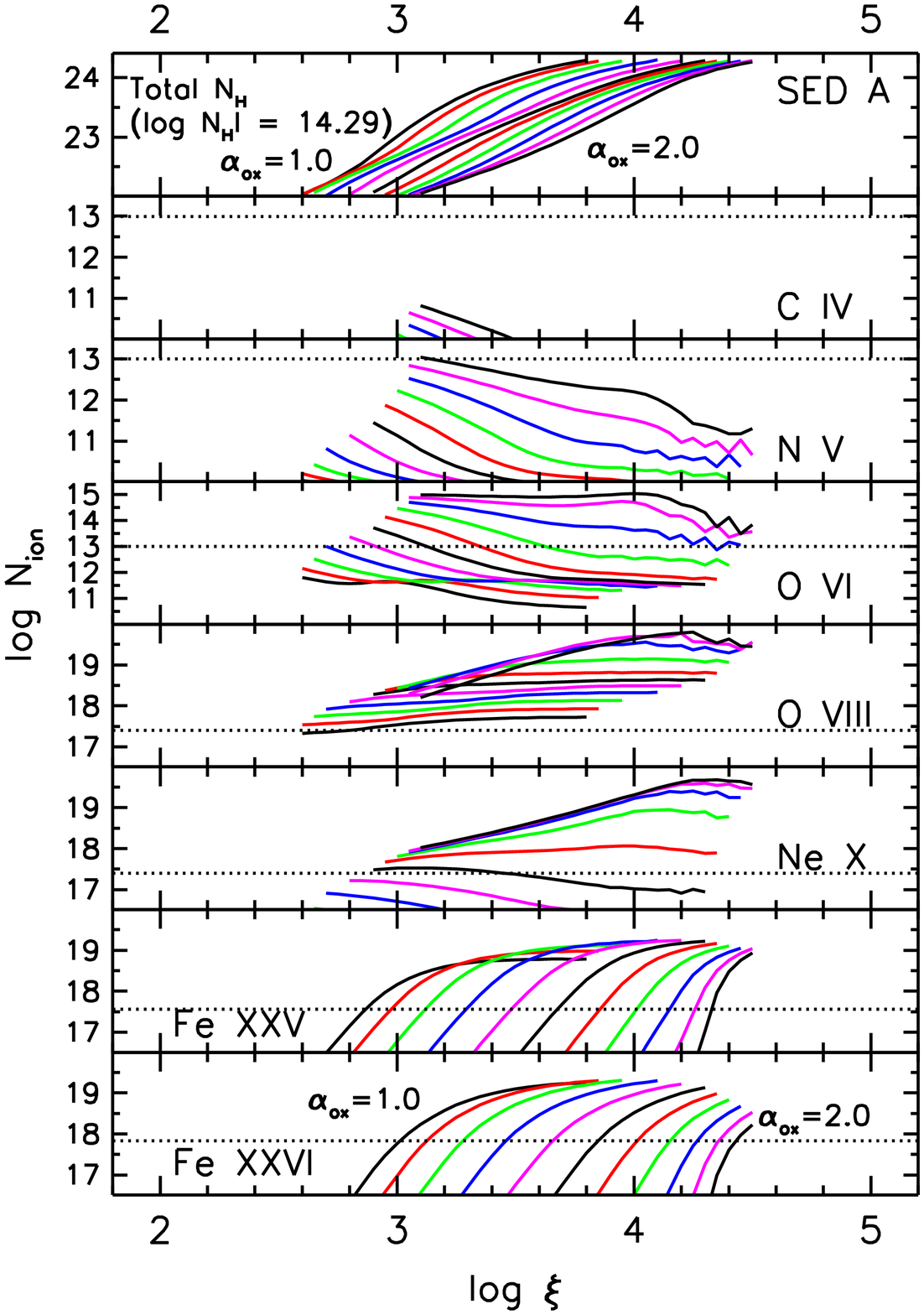}
\caption{
Ionic column densities as a function of ionization parameter for SED A
that correspond to an ionization model solution that produces a 1-\AA\ EW
Ly$\alpha$ absorption line. With a FWHM = $1000~\rm km~s^{-1}$, such a line
has a neutral hydrogen column density of log $\rm N_{HI} = 14.29~\rm cm^{-2}$.
The alternating color curves span a range in $\alpha_{ox}$ from 1.0 to
2.0 at intervals of 0.1.
Dashed horizontal lines show approximate thresholds for each ion for detecting
UV or X-ray absorption lines (see text).
}
\label{FigNtotA}
\end{figurehere}

\begin{figurehere}
\centering
\includegraphics[width=0.52\textwidth]{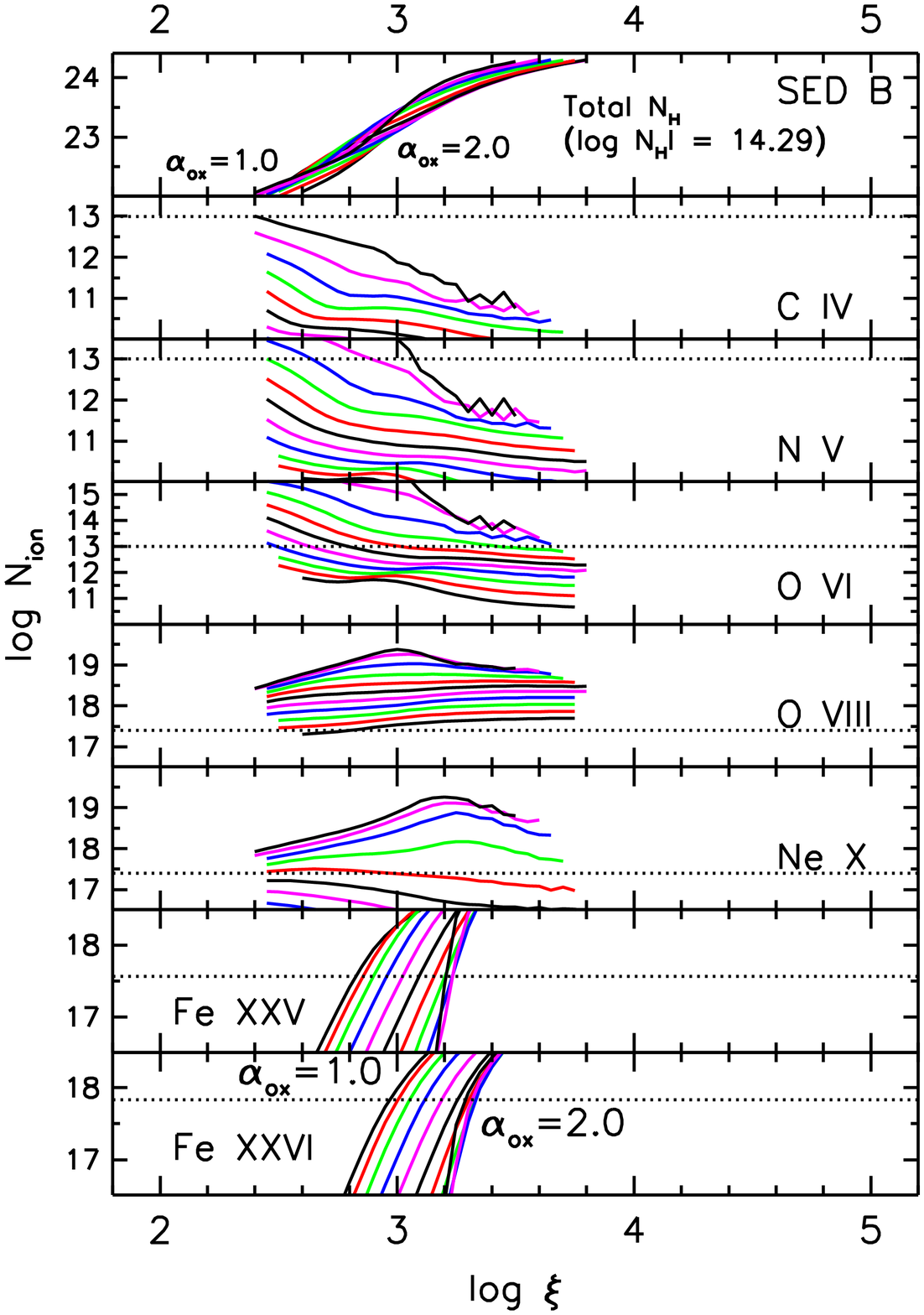}
\caption{
Ionic column densities as a function of ionization parameter for SED B
that correspond to an ionization model solution that produces a 1-\AA\ EW
Ly$\alpha$ absorption line. With a FWHM = $1000~\rm km~s^{-1}$, such a line
has a neutral hydrogen column density of log $\rm N_{HI} = 14.29~\rm cm^{-2}$.
The alternating color curves span a range in $\alpha_{ox}$ from 1.0 to
2.0 at intervals of 0.1.
Dashed horizontal lines show approximate thresholds for each ion for detecting
UV or X-ray absorption lines (see text).
}
\label{FigNtotB}
\end{figurehere}
\vskip 12pt

If we look at predicted column densities for an ionization parameter
log $\xi$ = 3.0 and $\alpha_{ox} = 1.5$, comparable to the conditions in
PG 1211+143, one sees that the total column density is
log $\rm N_{H} \sim 10^{22}~\rm cm^{-2}$, \ion{C}{4} and \ion{N}{5} are
invisible and \ion{O}{6} is possibly detectable at a
few$\times 10^{13}~\rm cm^{-2}$.
In the X-ray, \ion{O}{8} and \ion{Ne}{10} should be seen in a good grating spectrum, but the ionization level is too low to produce detectable
\ion{Fe}{25} or \ion{Fe}{26} features.

At slightly higher ionization parameters, \ion{C}{4} and \ion{N}{5} remain
undetectable, and \ion{O}{6} is strong only for soft spectra with high
$\alpha_{ox}$. However, \ion{O}{8} and \ion{Ne}{10} increase in strength,
and they should produce easily identifiable features in X-ray grating spectra.
\ion{Fe}{25} and \ion{Fe}{26} do not become detectable until log $\xi > 3.6$.
So, in the case of a strong UV \ion{H}{1} Ly$\alpha$ feature in which
\ion{Fe}{25} and \ion{Fe}{26} are also detected, one should also see
strong intermediate ionization ions such as \ion{O}{8} and \ion{Ne}{10}.
Again, we must conclude that strong \ion{H}{1} Ly$\alpha$ absorption should
be accompanied by other spectral features in addition to
\ion{Fe}{25} or \ion{Fe}{26}.

\begin{figure}
\centering
\includegraphics[angle=-90, width=0.52\textwidth]{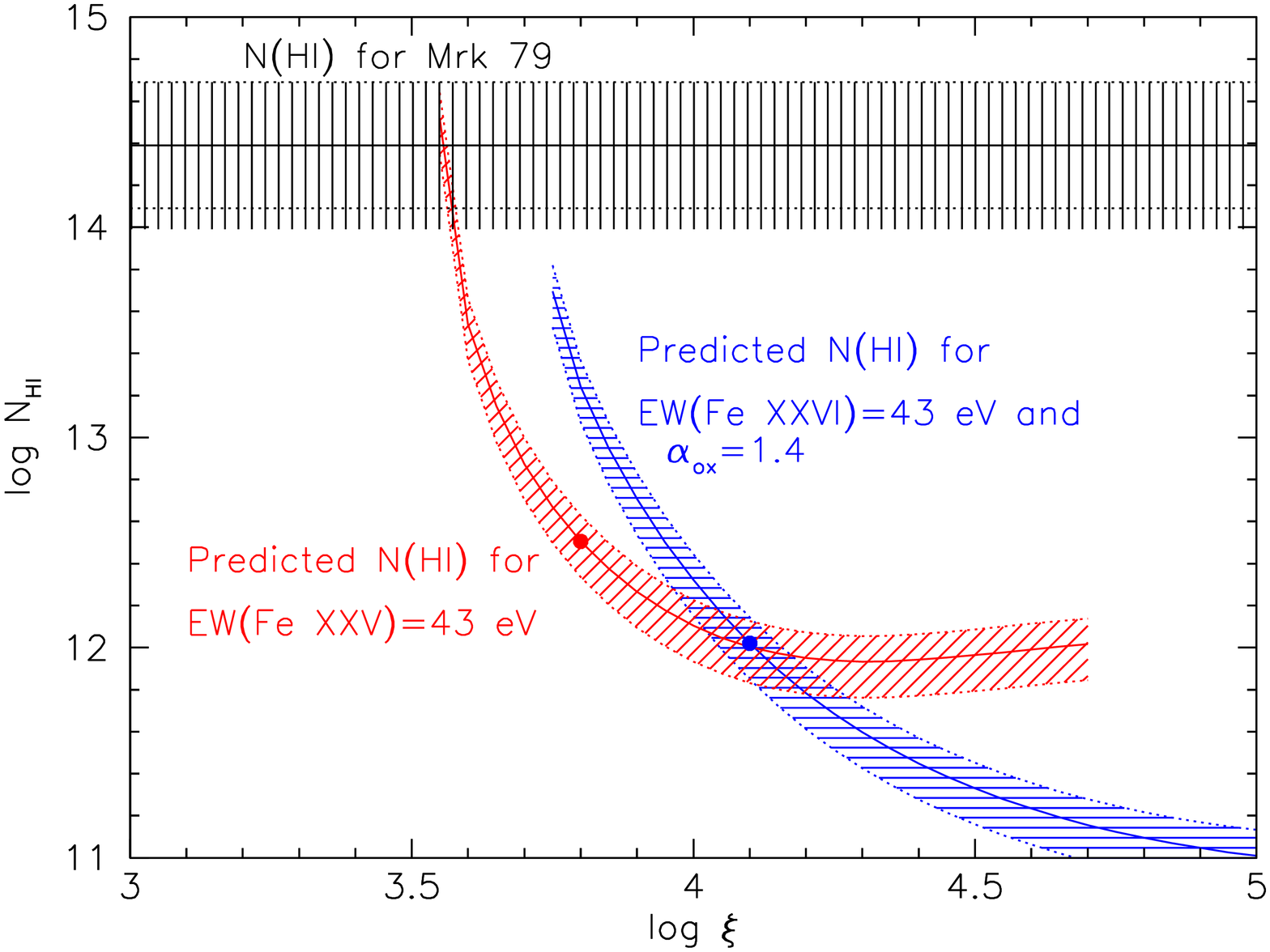}
\caption{
Predicted neutral hydrogen column densities for photoionization
models producing \ion{Fe}{25} or \ion{Fe}{26} absorption with EW=$43 \pm 14$ eV
in Mrk 79. The slanting curves show results using SED A for an
$\alpha_{ox} = 1.4$ with the shaded region showing the error bars corresponding
to the uncertainty in EW of the X-ray UFO.
The horizontal band shows the observed \ion{H}{1} column density with 1-$\sigma$
error bars.
The solid dots shows the locations of maximum ionization fraction for
\ion{Fe}{25} and \ion{Fe}{26} in the photoionization models.
}
\label{FigPredictedHI_Mrk79}
\end{figure}

In the more specific case of Mrk 79,
although our detection of Ly$\alpha$ is tentative,
evaluating its implications is instructive.
Since we know the overall SED of Mrk 79, we can more specifically
determine what ionization parameter will produce
both \ion{Fe}{26} and \ion{H}{1}, instead of examining a full
grid of possibilities as in Figure \ref{FigPredictedHIFe26}.
First we note that Figures \ref{FigPredictedHIFe26} and
\ref{FigPredictedHIFe25} show that the X-ray absorption feature
must be \ion{Fe}{25} in order to produce the observed column
density of neutral hydrogen.
Therefore, Figure \ref{FigPredictedHI_Mrk79}
shows curves similar to those in
Figure \ref{FigPredictedHIFe26}, but specifically computed for
EW=$43 \pm 18$ eV as reported by \cite{Tombesi10} with $\alpha_{ox}$=1.4
and SED A. The enclosed area then intersects the observed \ion{H}{1}
column density error region at only at the extreme lower limits for the
ionization parameter, log $\xi = 3.55$, significantly
below peak ion fraction for \ion{Fe}{25}, and well below the peak for
\ion{Fe}{26}.
The implied total hydrogen column for this model has
$\rm log~N_H = 4.2 \times 10^{23}~cm^{-2}$.
This case illustrates that detectable \ion{H}{1} is possible if the
Fe ionization fraction is considerably below peak.
\cite{Tombesi11} give log $\xi = 4.19 \pm 0.19$ and
$\rm N_H = (1.94 \pm 1.2) \times 10^{23}~cm^{-2}$,
which is a higher ionization parameter than necessary for producing
detectable \ion{H}{1}, but we also note that
the SED used in their photoionization model has an
extraordinarily high X-ray to optical luminosity ratio,
with $\alpha_{ox} = 1$.
However, our tentative solution of very high column density and lower ionization
parameter is problematic since it lies at such an extreme location in
parameter space.
In addition, as shown in Figure \ref{FigNtotA}, lower-ionization X-ray
species such as \ion{O}{8} and \ion{Ne}{10} should also be present, and they
are not.
If the UFO was present at the time of the UV observations,
it looked significantly different
in total column density and ionization than as reported in \cite{Tombesi11}.

Variability is a potential explanation for these discrepancies.
\cite{Gallo11} show that Mrk 79 exhibits significant variations in X-ray
flux both in the hard band (3--10 keV), where variations of factors of
2--3 are common, and up to ten-fold variations in flux at low energy, $< 1$ keV.
The UFO reported by \cite{Tombesi10} corresponds to a high flux state
of Mrk 79.
The {\it FUSE} observations of Mrk 79 precede the first two {\it XMM-Newton}
observations by 10--14 months. Both of these X-ray observations are in
intermediate to low flux states, factors of 2--10 lower.
If these states were comparable to the the one at the time of the
{\it FUSE} observations, a low ionization parameter consistent with our
possible detection of broad \ion{H}{1} would be possible.
However, this still leaves open the question why an ultra-fast outflow in
intermediate-ionization ions was not observed in a similar low-flux state
in 2008. Variability in column density is a possibility, but it may also be
true that broad, $\sim 1000~\rm km~s^{-1}$ width features would have been
hard to detect in the low-flux {\it XMM-Newton} RGS spectrum.

\vskip 12pt

Producing high column densities of \ion{H}{1} associated with \ion{Fe}{26}
UFOs becomes even more difficult for Ly$\alpha$ absorption features even
stronger than the possible feature in Mrk 79 or PG 1211+143.
To produce \ion{H}{1} column densities above log $\rm N_H > 15.35 ~cm^{-2}$,
as \cite{Hamann18} investigated as a possibility in PDS 456, ionization
parameters must be low and total column densities must be high.
This leads to strong ionization fronts in ions well below
the ionization of \ion{Fe}{25} and \ion{Fe}{26}, and all the intermediate
ionization species we have been investigating
(\ion{C}{4}, \ion{N}{5}, \ion{O}{6}, \ion{O}{8}, and \ion{Ne}{10})
produce strong spectral features.
\cite{Hamann18} draw the same conclusion, and rule out Ly$\alpha$ absorption
at $v=-0.06c~\rm km~s^{-1}$ as a possible identification for the
spectral feature in the 2000 STIS spectrum of PDS 456.

Our evaluation of the non-detection of Ly$\alpha$ counterparts to
highly ionized Fe X-ray absorption lines has assumed that the gas
producing the Fe absorption as well as the Ly$\alpha$ absorption arises
in the same parcel of photoionized gas.
However, several authors have suggested that lower-ionization gas might be
embedded as denser clumps in the more highly ionized Fe outflows
\citep{Longinotti15,Hagino2017,Hamann18}.
In such models, one might expect Ly$\alpha$ absorption to be even more common,
as denser, lower-ionization clumps embedded in the UFO would have a greater
propensity to produce higher column densities in lower-ionization species
such as \ion{H}{1} (and \ion{C}{4}, \ion{N}{5}, and \ion{O}{6}).
Given that we do not detect \ion{H}{1} absorption in most cases, it seems
likely that co-existing clumps of denser, lower ionization gas, embedded in
UFOs is also not a common phenomenon.
However, mitigating possibilities to this conclusion would be that the
entrained, high-density material does not have the same velocity, or, as
we discuss below, variability.

Variability is a confounding factor in interpreting these results.
As \cite{Kriss18} showed, the broad Ly$\alpha$ absorption detected
in PG 1211+143 is not always present.
\cite{Reeves18} also see dramatic variability in the soft X-ray
spectrum of the $v_{out} \sim -18\,000~\rm km~s^{-1}$ outflow in
PG 1211+143. At the time of their {\it XMM-Newton} observation
it was not only brighter, with higher ionization parameter
and higher total column density than that observed by
\cite{Danehkar18} when the Ly$\alpha$ absorption was present,
but it also had a strong, low-ionization zone with
log $\xi$ = 1.32 and $\rm N_H = 1.3 \times 10^{21}~\rm cm^{-2}$.
This gas would have produced strong, deep UV lines in
\ion{H}{1}, \ion{C}{4}, \ion{N}{5}, and \ion{O}{6}, all of which
would have been easily detected in the HST/COS observations of
\cite{Danehkar18} and \cite{Kriss18}.
Such dramatic variability means that non-detections and
upper limits derived from non-simultaneous observations
can represent conditions that differ by orders of magnitude
in column density of the outflowing gas.
More readily interpreted results require simultaneous
UV and X-ray spectra.
\vskip 12pt

\bigskip
\bigskip
\section{Summary}

We have examined 36 archival UV spectra to search for potential
\ion{H}{1} Ly$\alpha$ absorption counterparts to a sample of
16 X-ray ultra-fast outflows selected from \cite{Tombesi10}
and \cite{Gofford13}.
Two FUSE spectra of Mrk 79 show evidence for broad Ly$\alpha$
absorption at the expected wavelength for its UFO, but
these detections are tentative since they appear in only one
of two simultaneous exposures.
This tentative feature has FWHM=910 $\rm km~s^{-1}$,
an equivalent width of 1.3 \AA, and an implied
\ion{H}{1} column density of $2.5 \times 10^{14}~\rm cm^{-2}$.
If the X-ray UFO has an ionization parameter of log $\xi$=3.55
and a total column density of $\rm N_H = 4.2 \times 10^{23}~cm^{-2}$,
then the same gas can produce both the \ion{H}{1} and the UFO, but the
X-ray feature must be \ion{Fe}{25} and not \ion{Fe}{26}.
One would also expect to detect lower-ionization species
(e.g., \ion{O}{8}) in the X-ray spectrum of Mrk 79 at the UFO
velocity, and they are not present.

Applying a search criterion of a line width of 1000 $\rm km~s^{-1}$,
comparable to the Ly$\alpha$ counterpart to an X-ray UFO in
PG 1211+143 discovered by \cite{Kriss18}, we can set upper limits
on associated \ion{H}{1} column densities in the range from
log $N_{HI}$=12.7 to 14.4 $\rm cm^{-2}$, with most
having log $N_{HI} < 13.3~\rm cm^{-2}$.
We have computed a set of photoionization models covering a
large grid of possible SEDs that shows that if most
X-ray UFO absorbers are \ion{Fe}{25} or \ion{Fe}{26} near
their peak ionization fractions, the associated column
densities of \ion{H}{1} would be lower than these limits.
In general, ionization parameters have to be higher than
log $\xi$=3.7 to make \ion{H}{1} undetectable.
Lower ionization parameters can produce detectable UV
absorbing gas, however, so an alternative explanation of
the paucity of counterparts in our search is that variability
only renders them visible on rare occasions, or that the
UV ionizing continuum is very hard.

\acknowledgments
We thank M. Cappi for suggesting this study.
All of the data presented in this paper were obtained from the
Mikulski Archive for Space Telescopes (MAST).
STScI is operated by the Association of Universities for Research in Astronomy,
Inc., under NASA contract NAS5-26555.
Support for MAST for non-HST data is provided by the NASA Office of Space
Science via grant NNX09AF08G and by other grants and contracts.
This work was supported by NASA through grant NNX13AF18G, and by
a grant for {\it HST} program number 13233
from the Space Telescope Science Institute (STScI), which is
operated by the Association of Universities for Research
in Astronomy, Incorporated, under NASA contract NAS5-26555.
This research has made use of the NASA/IPAC Extragalactic Database (NED), 
which is operated by the 
Jet Propulsion Laboratory, California Institute of Technology, 
under contract with the National Aeronautics and Space Administration.

\bibliographystyle{apj}
\bibliography{limits}

\newpage
\setcounter{figure}{0} 
\begin{figure*}[htb]
\centering
  \begin{subfigure}[b]{0.48\linewidth}
    \centering
    \includegraphics[angle=-90,width=.99\textwidth]{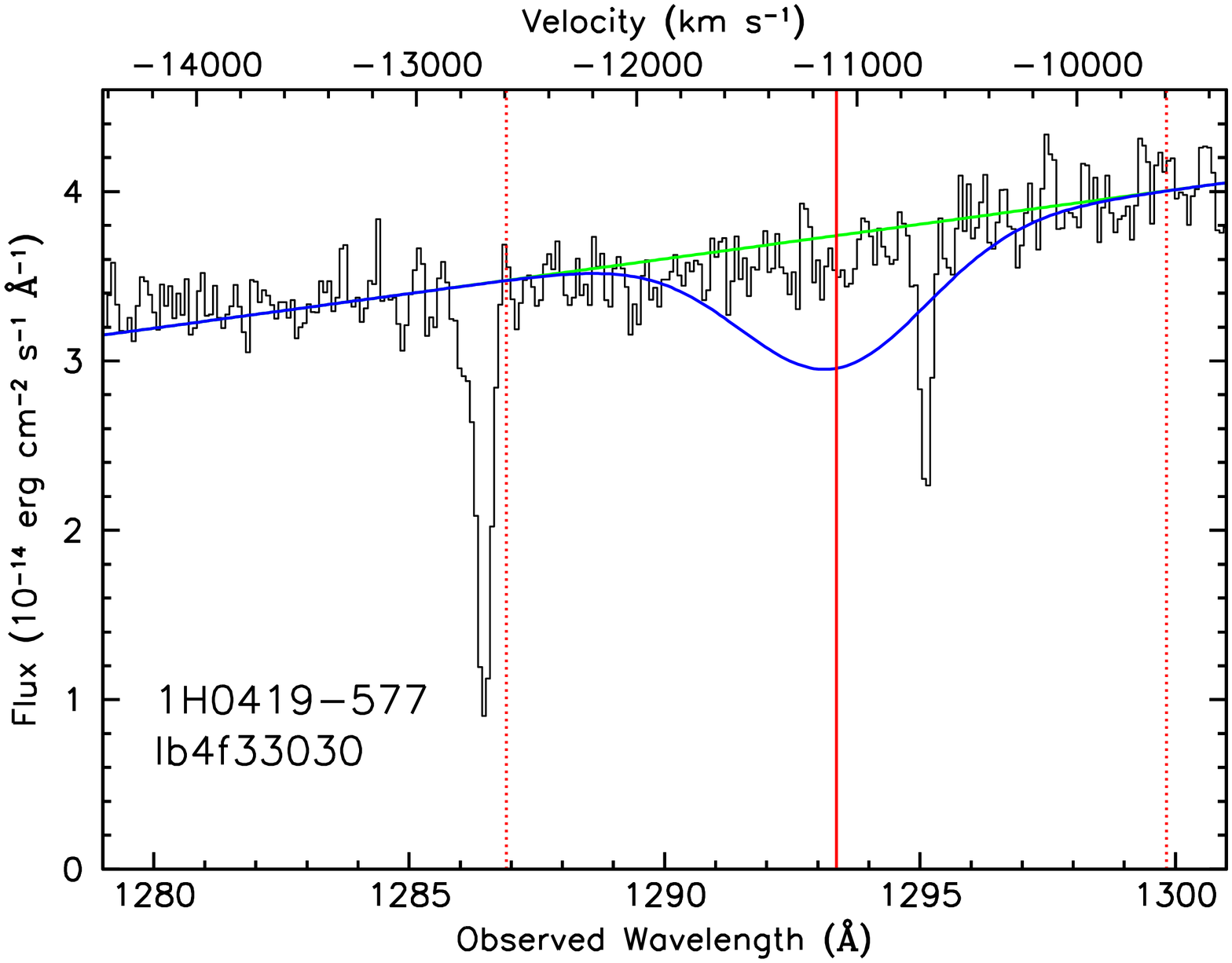}
    \caption{1H0419$-$577, HST observation lb4f03030}\label{fig:1a}
  \end{subfigure}%
  \begin{subfigure}[b]{0.48\linewidth}
    \centering
    \includegraphics[angle=-90,width=.99\textwidth]{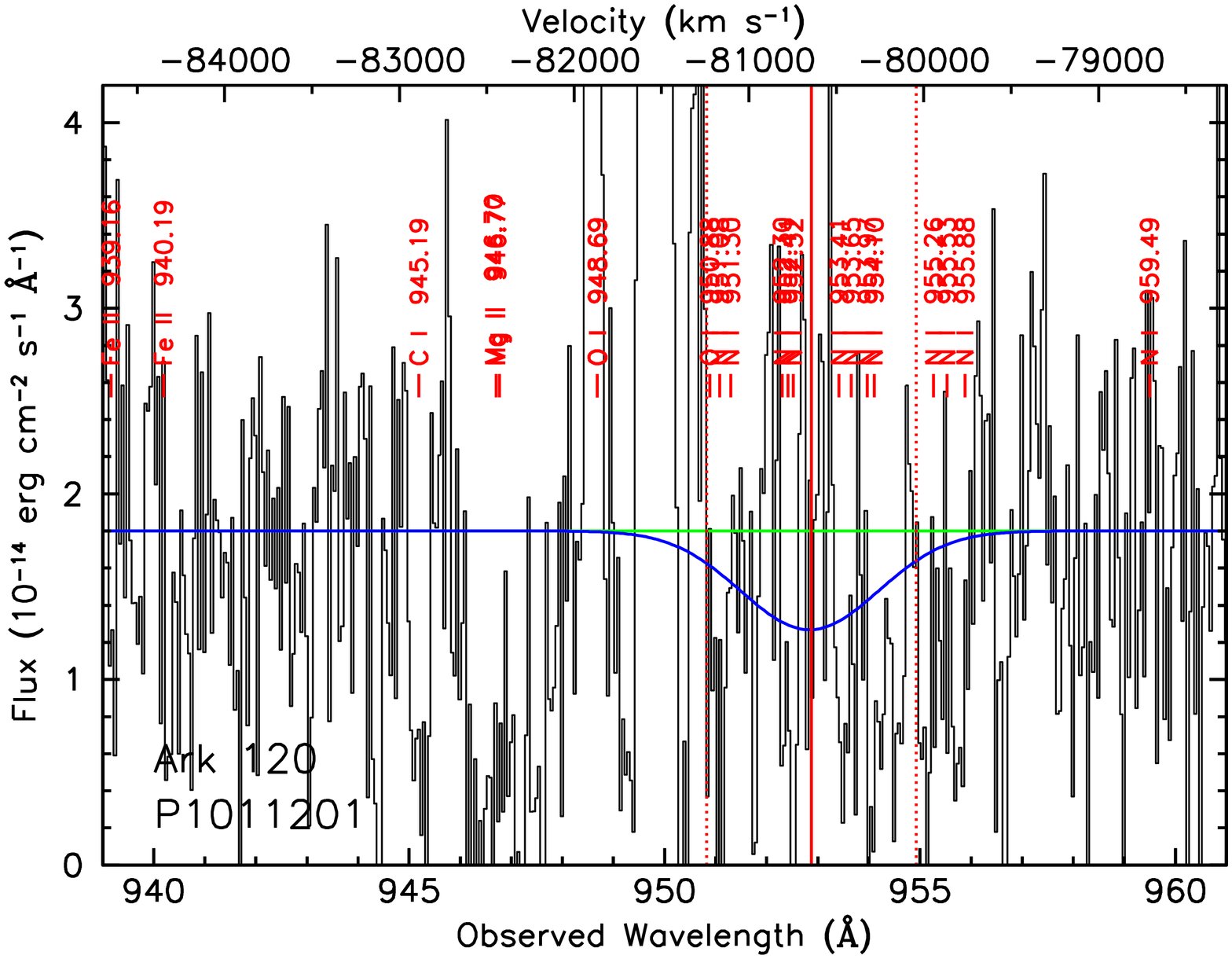}
    \caption{Ark 120, FUSE observation P1011201}\label{fig:1b}
  \end{subfigure}%
  \begin{subfigure}[b]{0.48\linewidth}
    \centering
    \includegraphics[angle=-90,width=.99\textwidth]{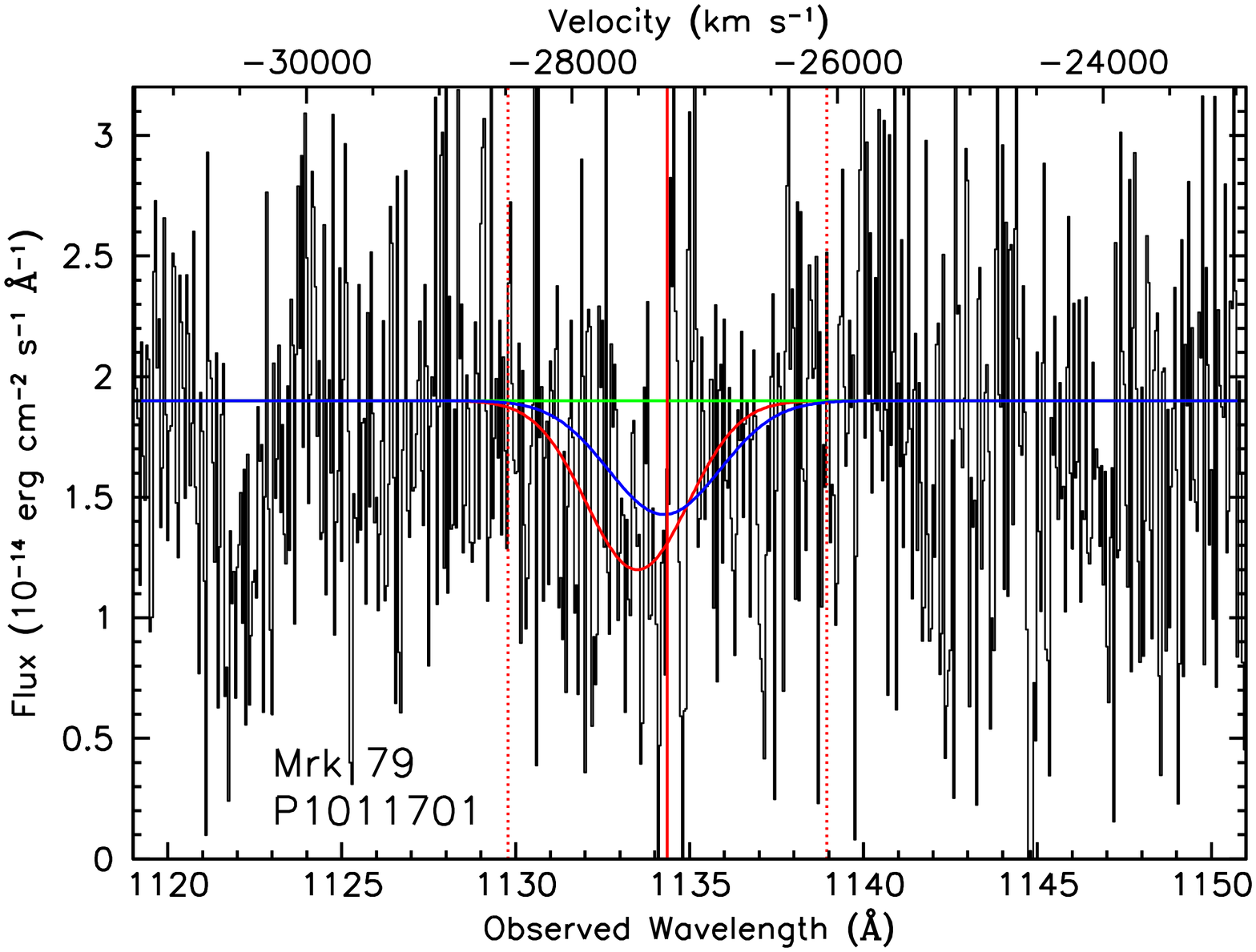}
    \caption{Mrk 79, FUSE observation P1011701}\label{fig:1c}
  \end{subfigure}%
  \begin{subfigure}[b]{0.48\linewidth}
    \centering
    \includegraphics[angle=-90,width=.99\textwidth]{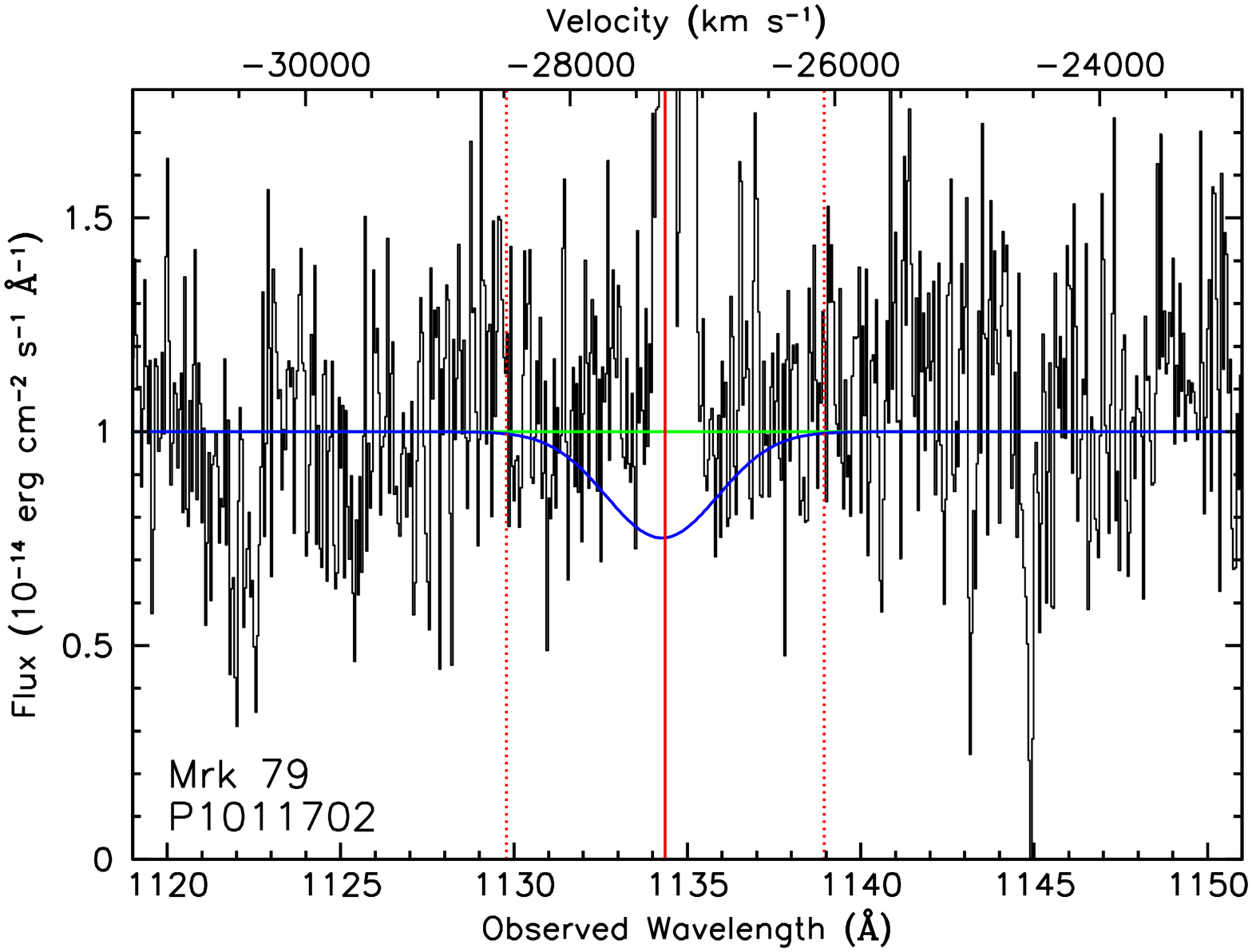}
    \caption{Mrk 79, FUSE observation P1011702}\label{fig:1d}
  \end{subfigure}%
  \begin{subfigure}[b]{0.48\linewidth}
    \centering
    \includegraphics[angle=-90,width=.99\textwidth]{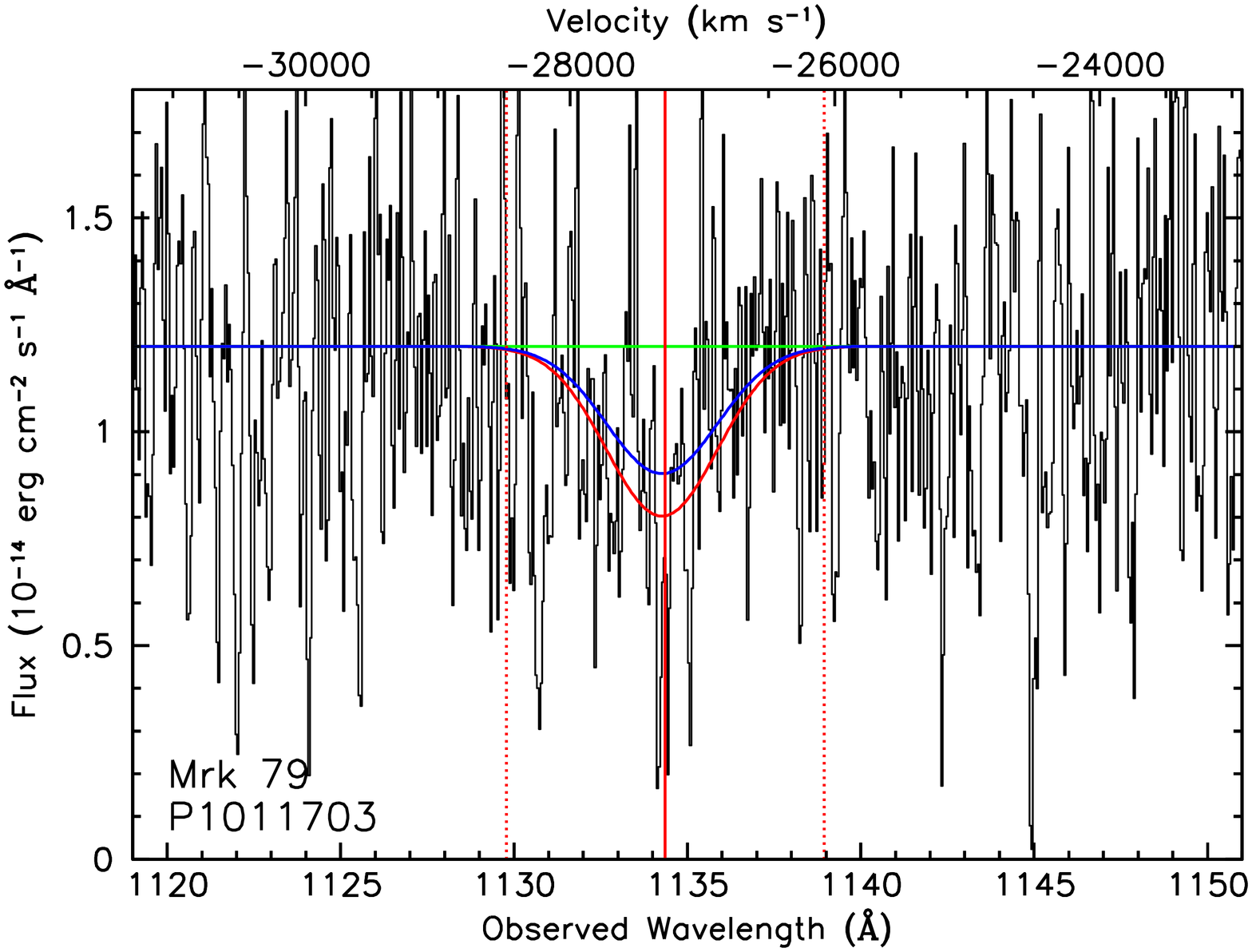}
    \caption{Mrk 79, FUSE observation P1011703}\label{fig:1e}
  \end{subfigure}%
  \begin{subfigure}[b]{0.48\linewidth}
    \centering
    \includegraphics[angle=-90,width=.99\textwidth]{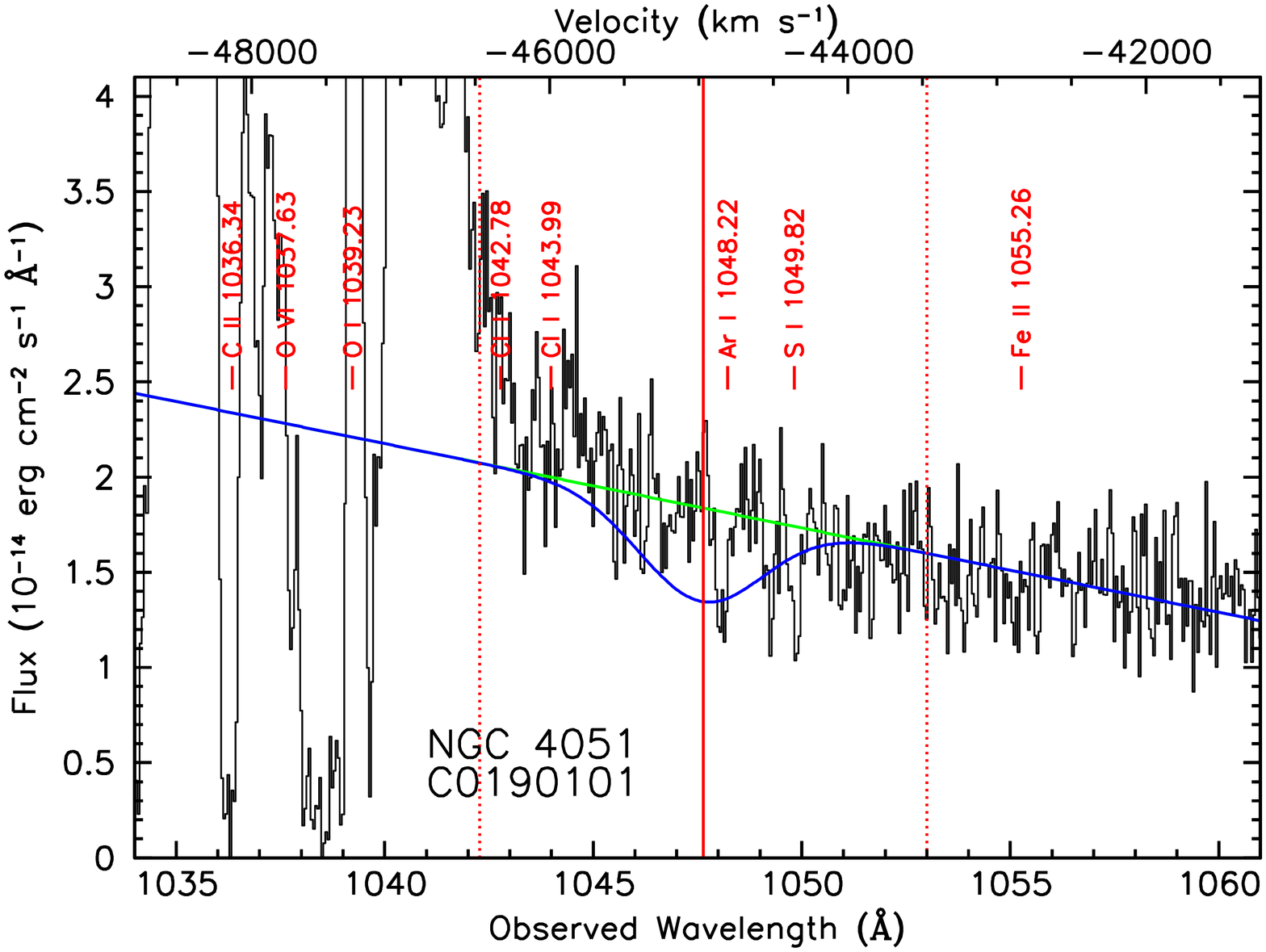}
    \caption{NGC 4051, FUSE observation C0190101}\label{fig:1f}
  \end{subfigure}\\%
  \caption{FUSE and HST spectra of regions surrounding the predicted positions
           of Ly$\alpha$ counterparts to X-ray UFOs.  A thin red vertical line
           marks the predicted wavelength, and the two red vertical dashed
           lines gives the error region for the prediction.
           The lower horizontal scale gives the observed wavelength.
           The upper horizontal scale gives the outflow velocity in $\rm km~s^{-1}$.
           The green line gives the fitted continuum at the location of predicted Ly$\alpha$.
           The blue line shows the best fit, incorporating additional emission and absorption
           components as needed for each object. In addition, the blue line illustrates the
           predicted appearance of a 1 \AA\ EW absorption line with FWHM=1000 $\rm km~s^{-1}$
           at the predicted position of Ly$\alpha$.  Red lines in Figures \ref{fig:1c} and \ref{fig:1e} for Mrk 79
           show the best fit for detected Ly$\alpha$ absorption lines.
	   For some spectra, interstellar absorption lines are labeled in red.}\label{FigFitsa}
\end{figure*}

\newpage
\begin{figure*}[htb]
\centering
  \addtocounter{figure}{-1} 
  \begin{subfigure}[b]{.40\linewidth}
    \centering
    \includegraphics[angle=-90,width=.99\textwidth]{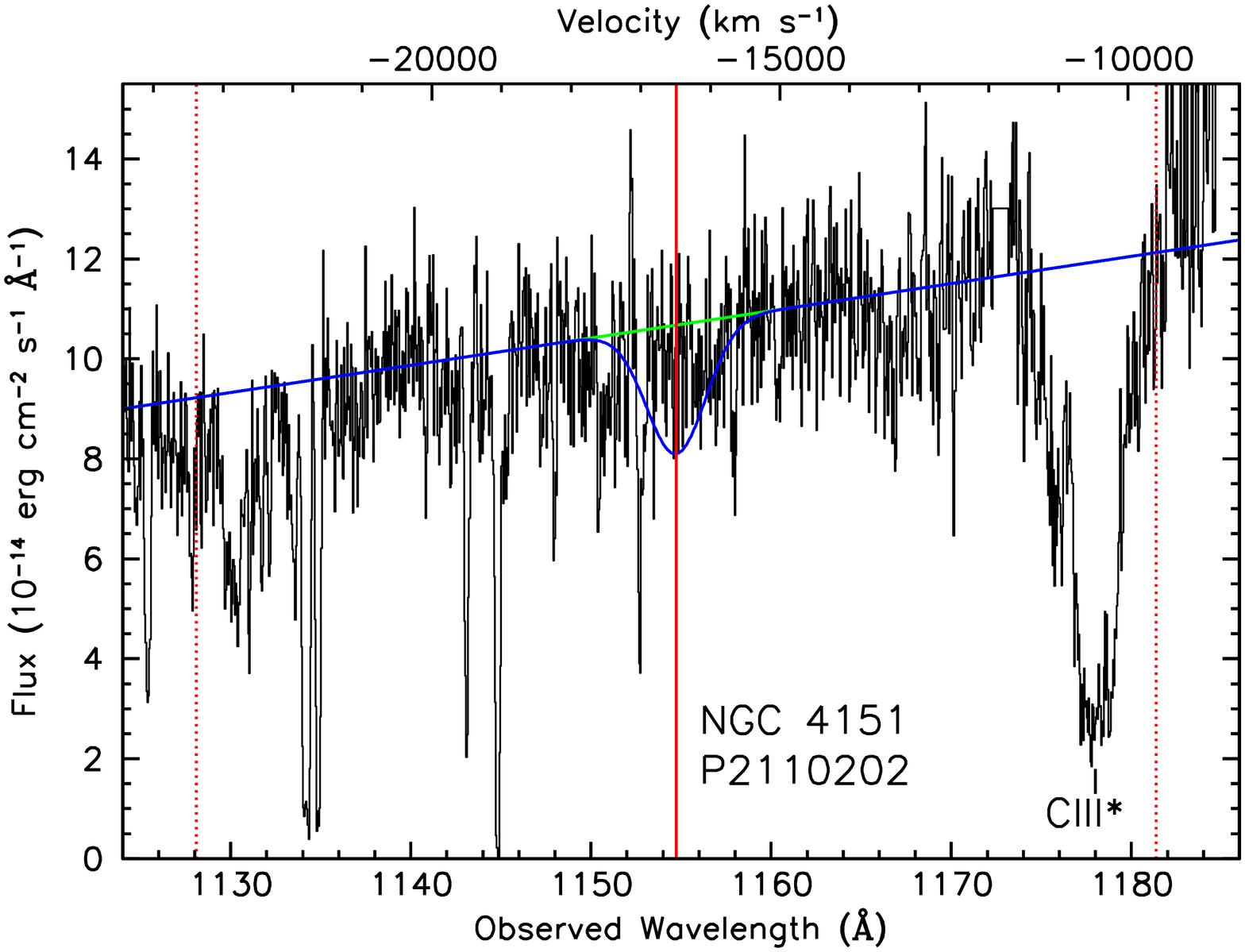}
    \addtocounter{subfigure}{6}
    \caption{NGC 4151, FUSE observation P2110202}\label{fig:1g}
  \end{subfigure}%
  \begin{subfigure}[b]{.40\linewidth}
    \centering
    \includegraphics[angle=-90,width=.99\textwidth]{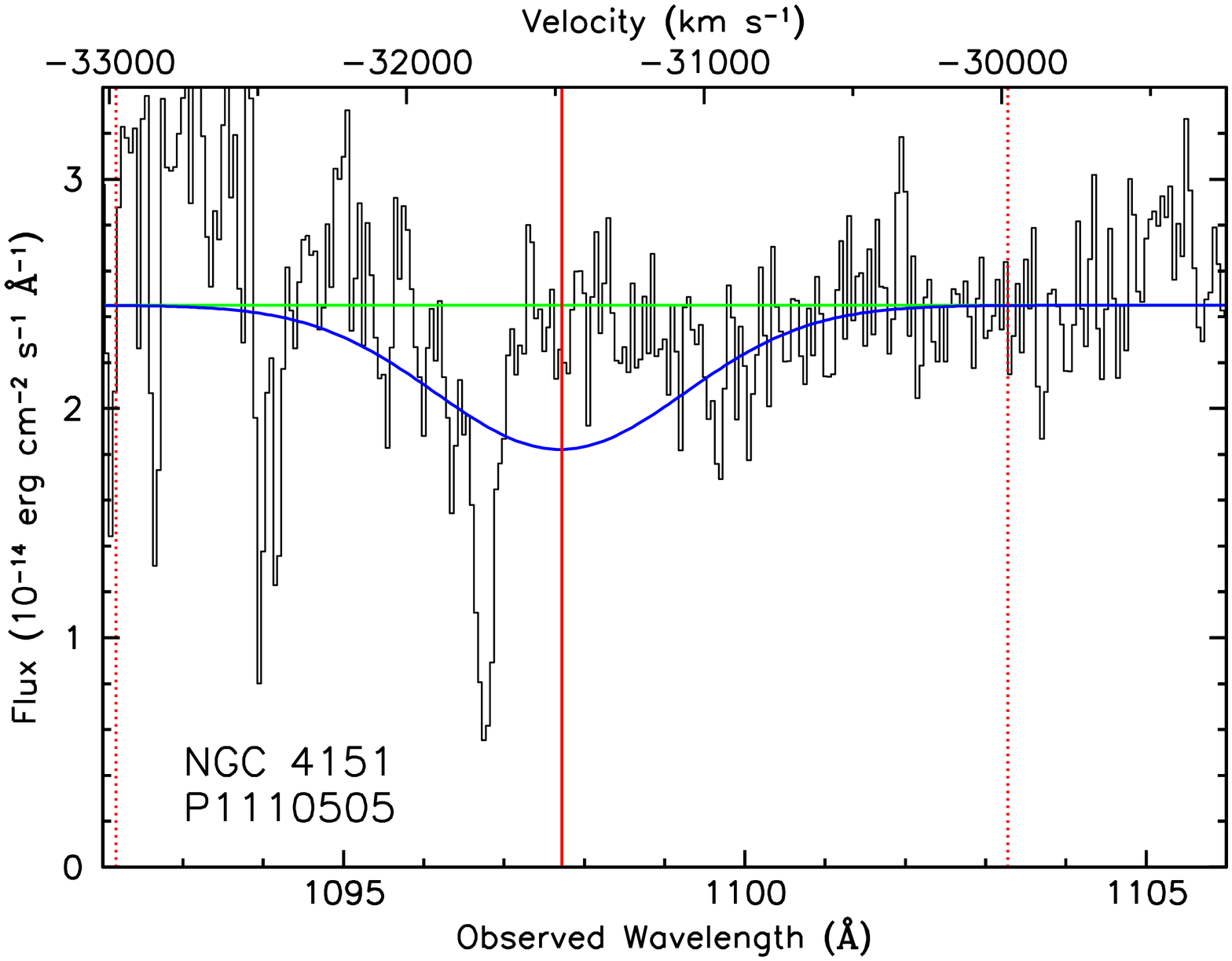}
    \caption{NGC 4151, FUSE observation P1110505}\label{fig:1h}
  \end{subfigure}%
  \begin{subfigure}[b]{.40\linewidth}
    \centering
    \includegraphics[angle=-90,width=.99\textwidth]{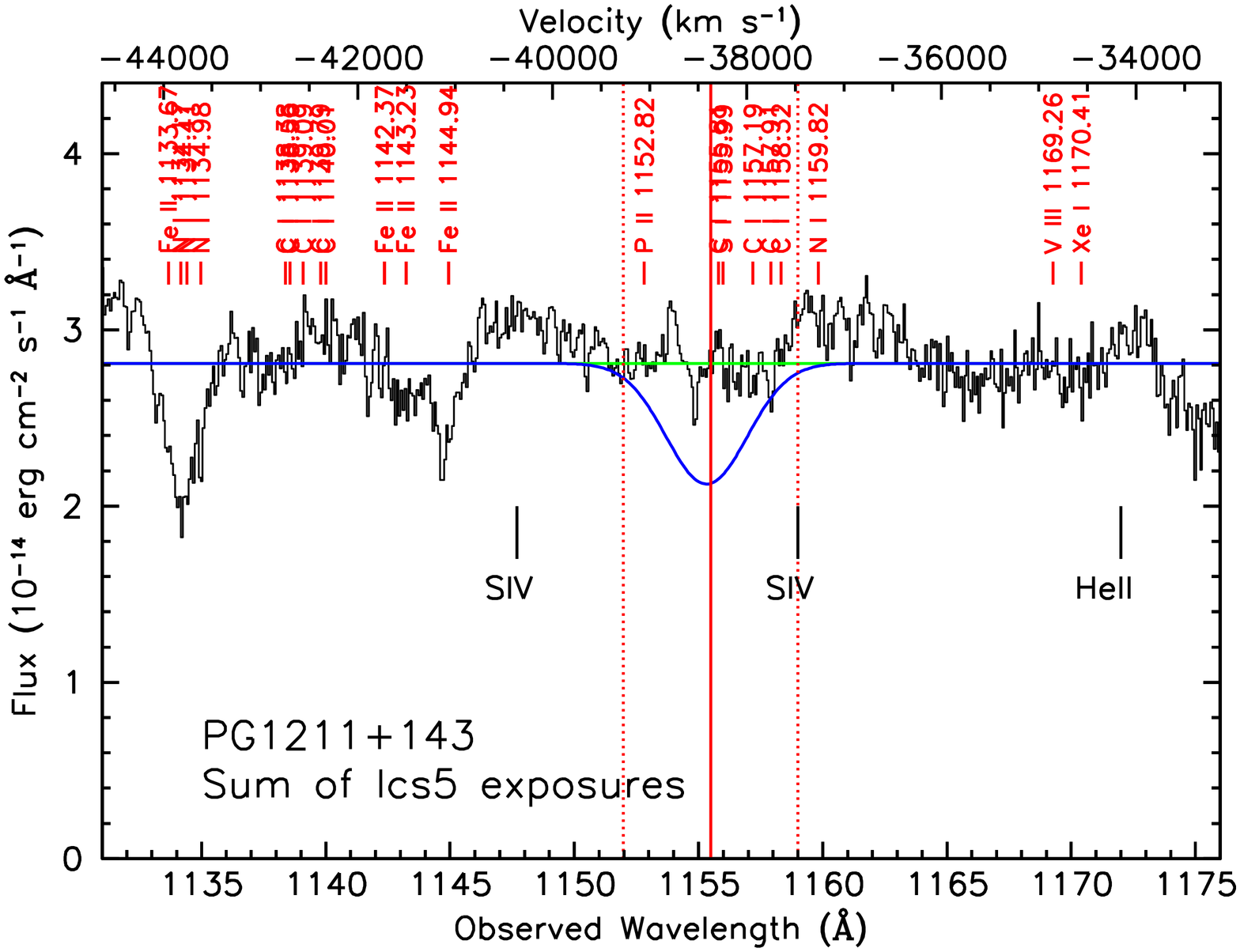}
    \caption{PG 1211+143, Sum of HST observations\\
lcs501010, lcs502010, and lcs504010}\label{fig:1i}
  \end{subfigure}%
  \begin{subfigure}[b]{.40\linewidth}
    \centering
    \includegraphics[angle=-90,width=.99\textwidth]{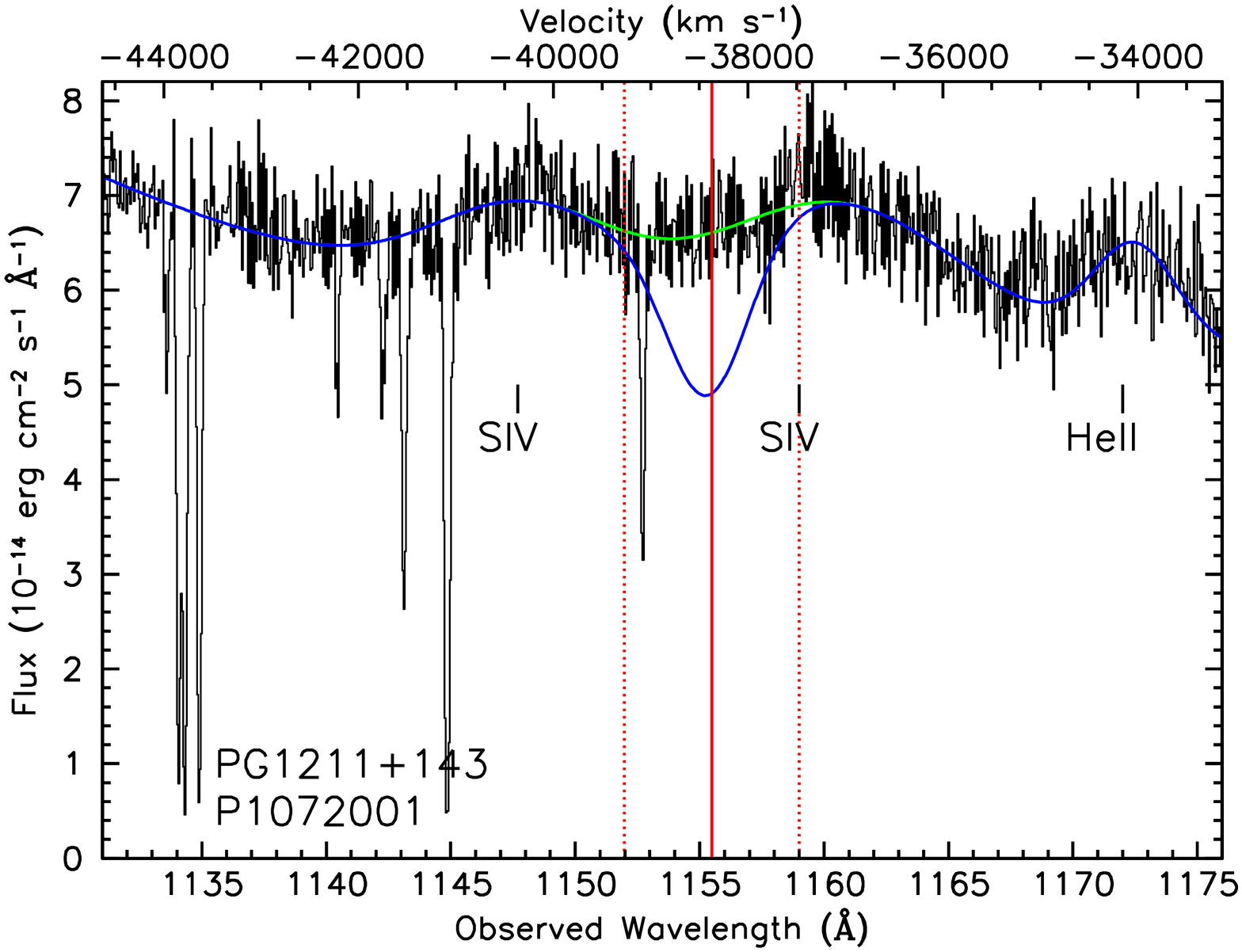}
    \caption{PG 1211+143, FUSE observation P1072001\\
   $\phantom{ Sum of HST observations lcs501010}$}\label{fig:1j}
  \end{subfigure}%
  \begin{subfigure}[b]{.40\linewidth}
    \centering
    \includegraphics[angle=-90,width=.99\textwidth]{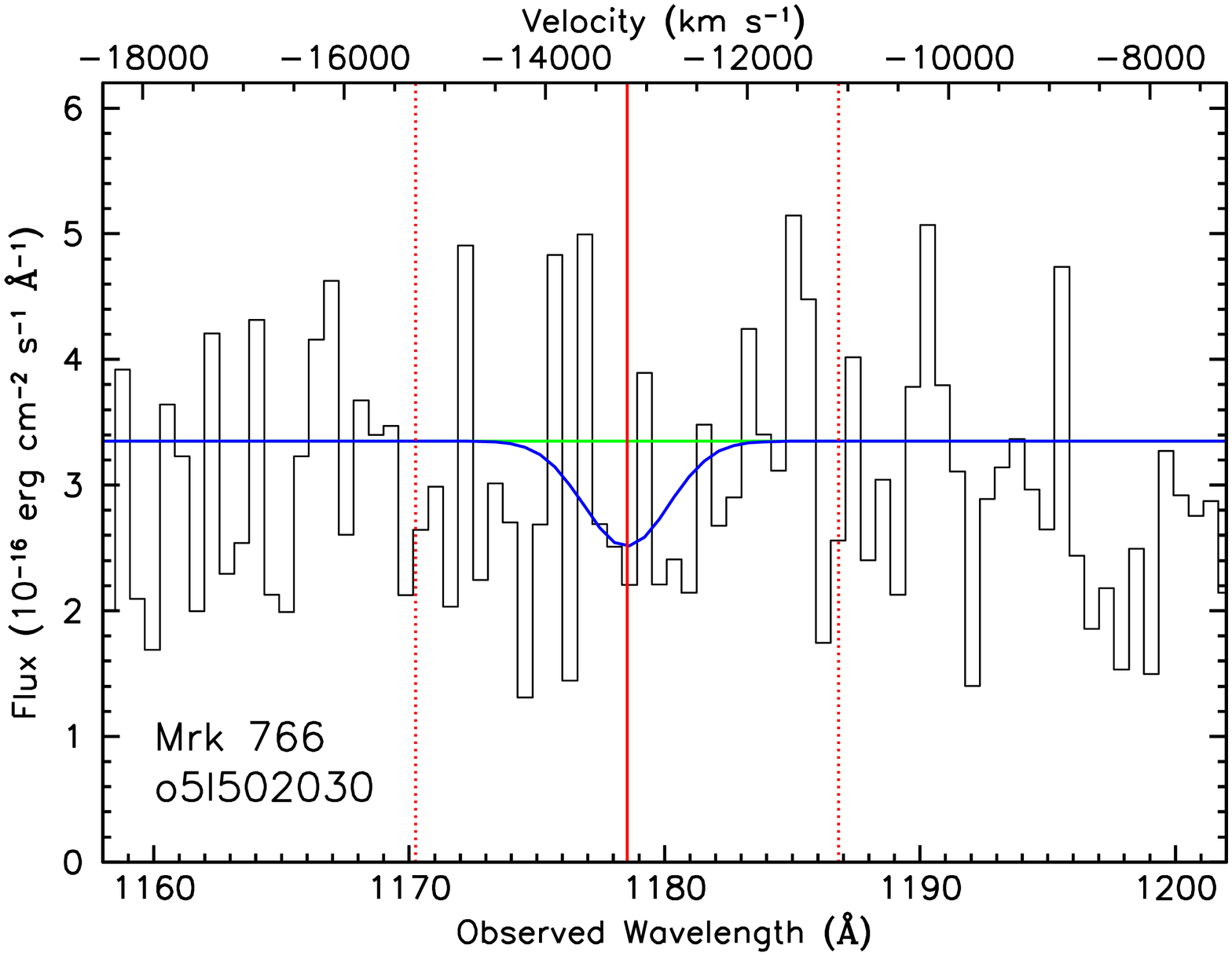}
    \caption{Mrk 766, HST observation o5l502030}\label{fig:1k}
  \end{subfigure}%
  \begin{subfigure}[b]{.40\linewidth}
    \centering
    \includegraphics[angle=-90,width=.99\textwidth]{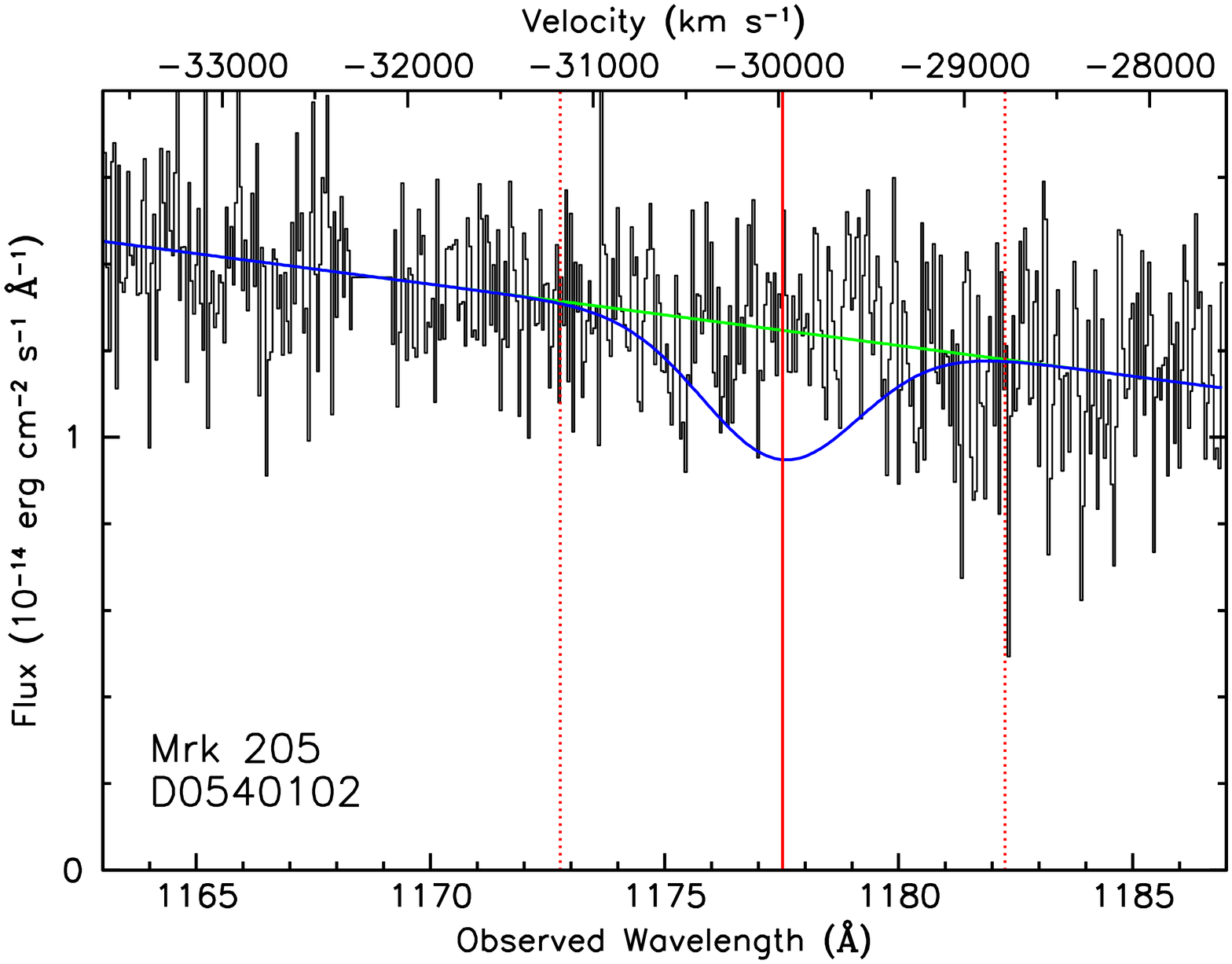}
    \caption{Mrk 205, FUSE observation D0540102}\label{fig:1l}
  \end{subfigure}%
  \begin{subfigure}[b]{.40\linewidth}
    \centering
    \includegraphics[angle=-90,width=.99\textwidth]{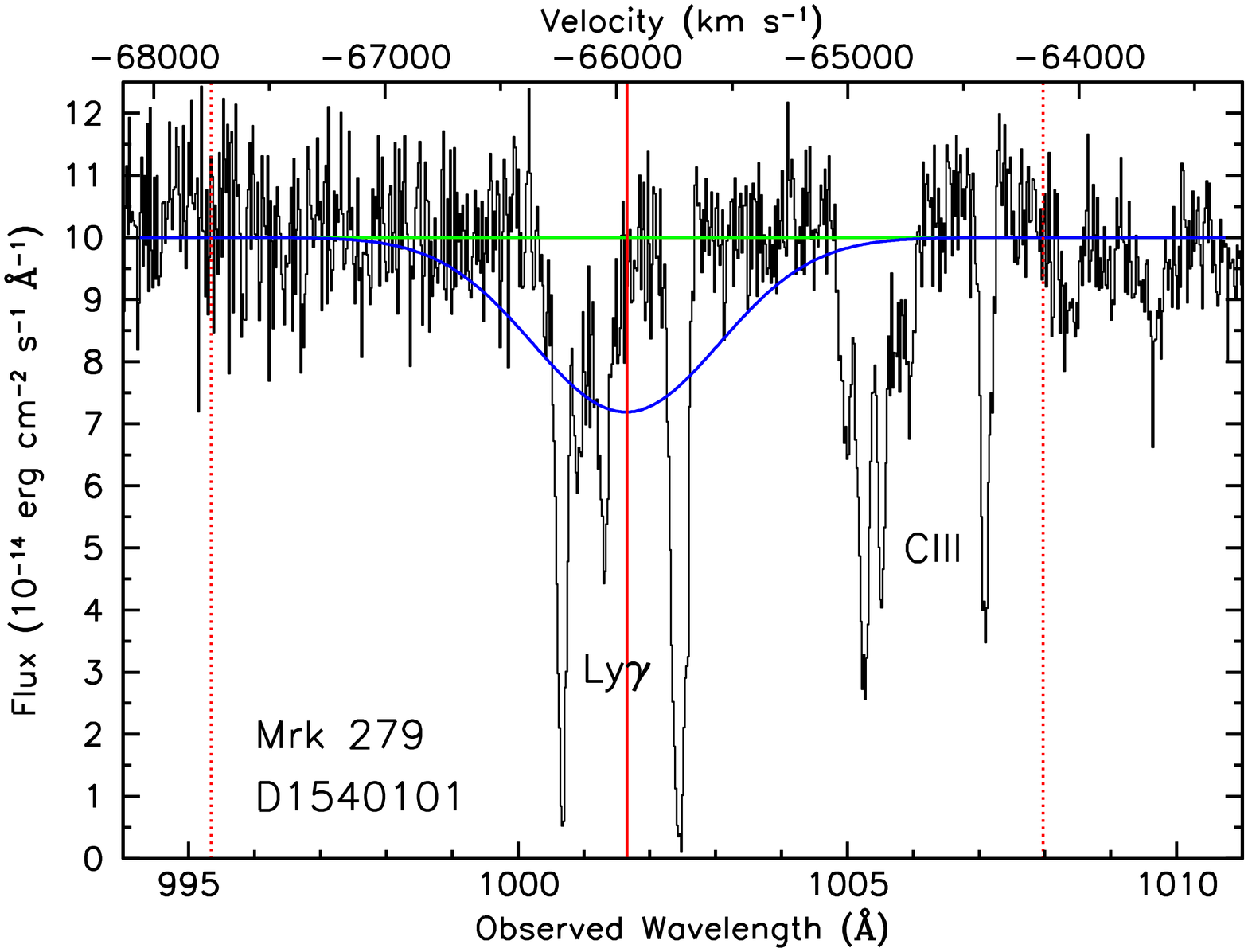}
    \caption{Mrk 279, FUSE observation D1540101. Intrinsic absorption in Ly$\gamma$ and \ion{C}{3} $\lambda$977 are marked.}\label{fig:1m}
  \end{subfigure}%
  \caption{(cont.)}\label{FigFitsb}
  \addtocounter{figure}{-1} 
\end{figure*}

\newpage
\begin{figure*}[htb]
\centering
  \begin{subfigure}[b]{.40\linewidth}
    \centering
    \includegraphics[angle=-90,width=.99\textwidth]{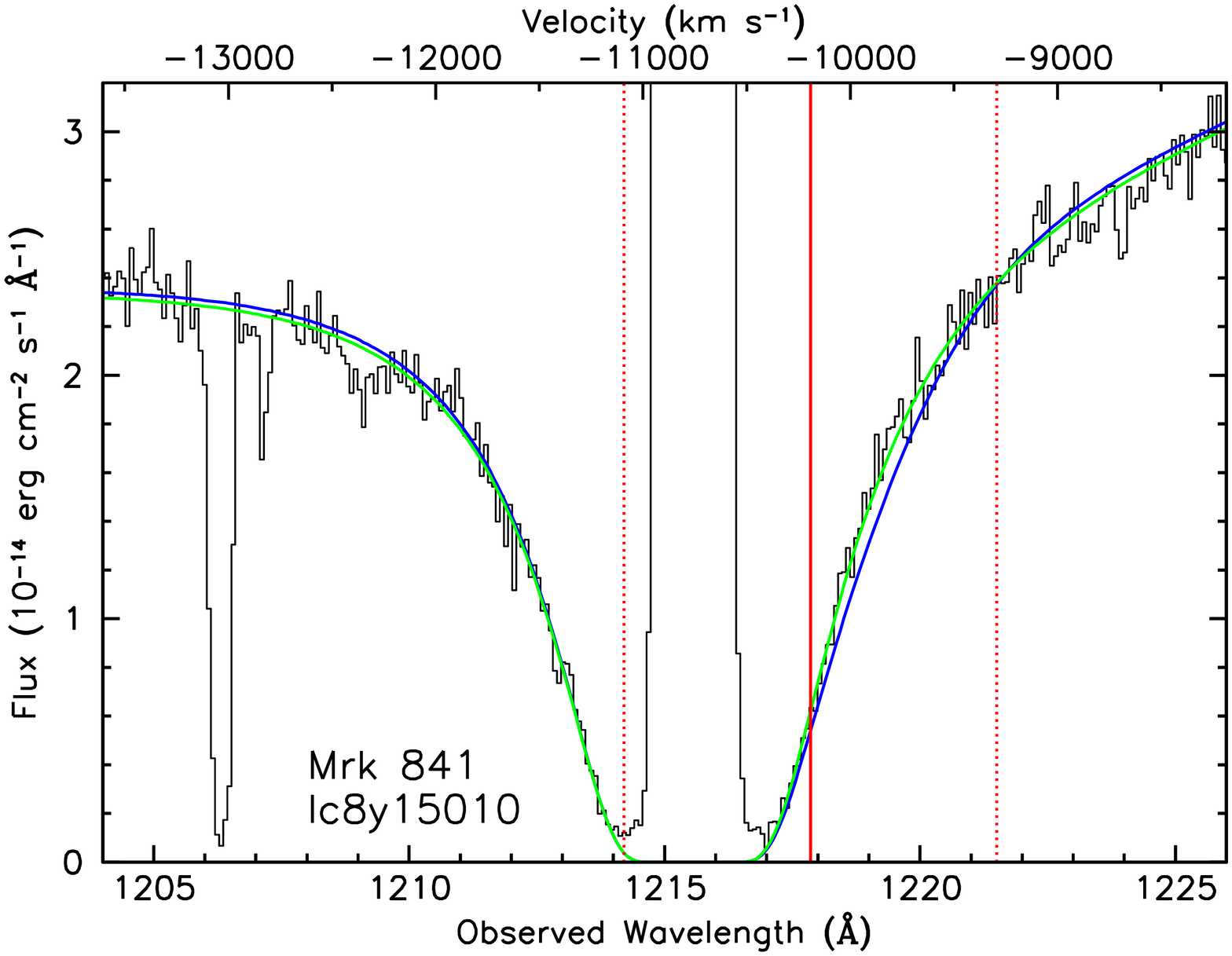}
    \addtocounter{subfigure}{13}
    \caption{Mrk 841, HST observation lc8y15010\\
   $\phantom{The red line shows the 2sigma upper limit.}$}\label{fig:1n}
  \end{subfigure}%
  \begin{subfigure}[b]{.40\linewidth}
    \centering
    \includegraphics[angle=-90,width=.99\textwidth]{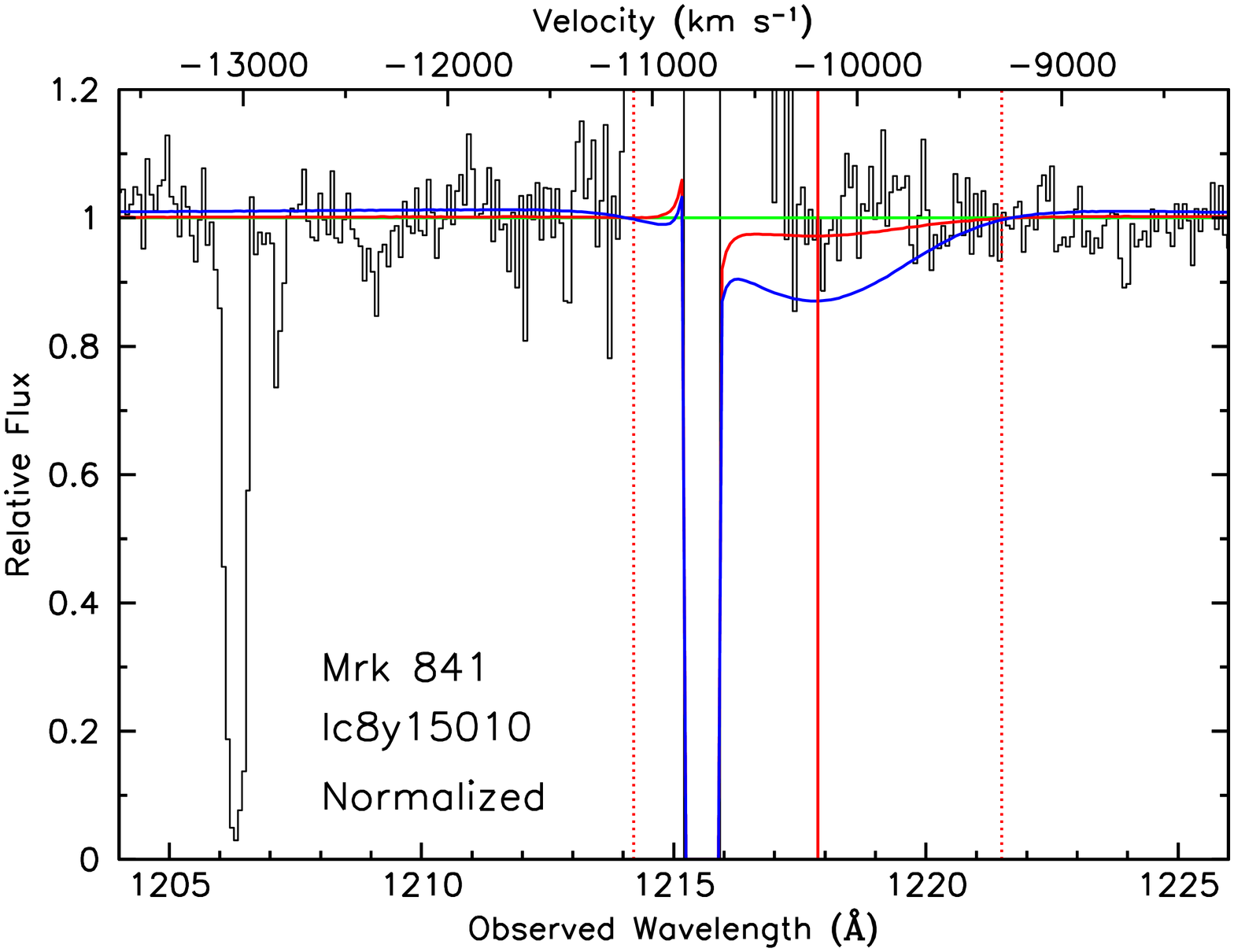}
    \caption{Mrk 841, normalized best-fit spectrum. The red line shows the 2$\sigma$ upper limit.}\label{fig:1o}
  \end{subfigure}\\%
  \begin{subfigure}[b]{.40\linewidth}
    \centering
    \includegraphics[angle=-90,width=.99\textwidth]{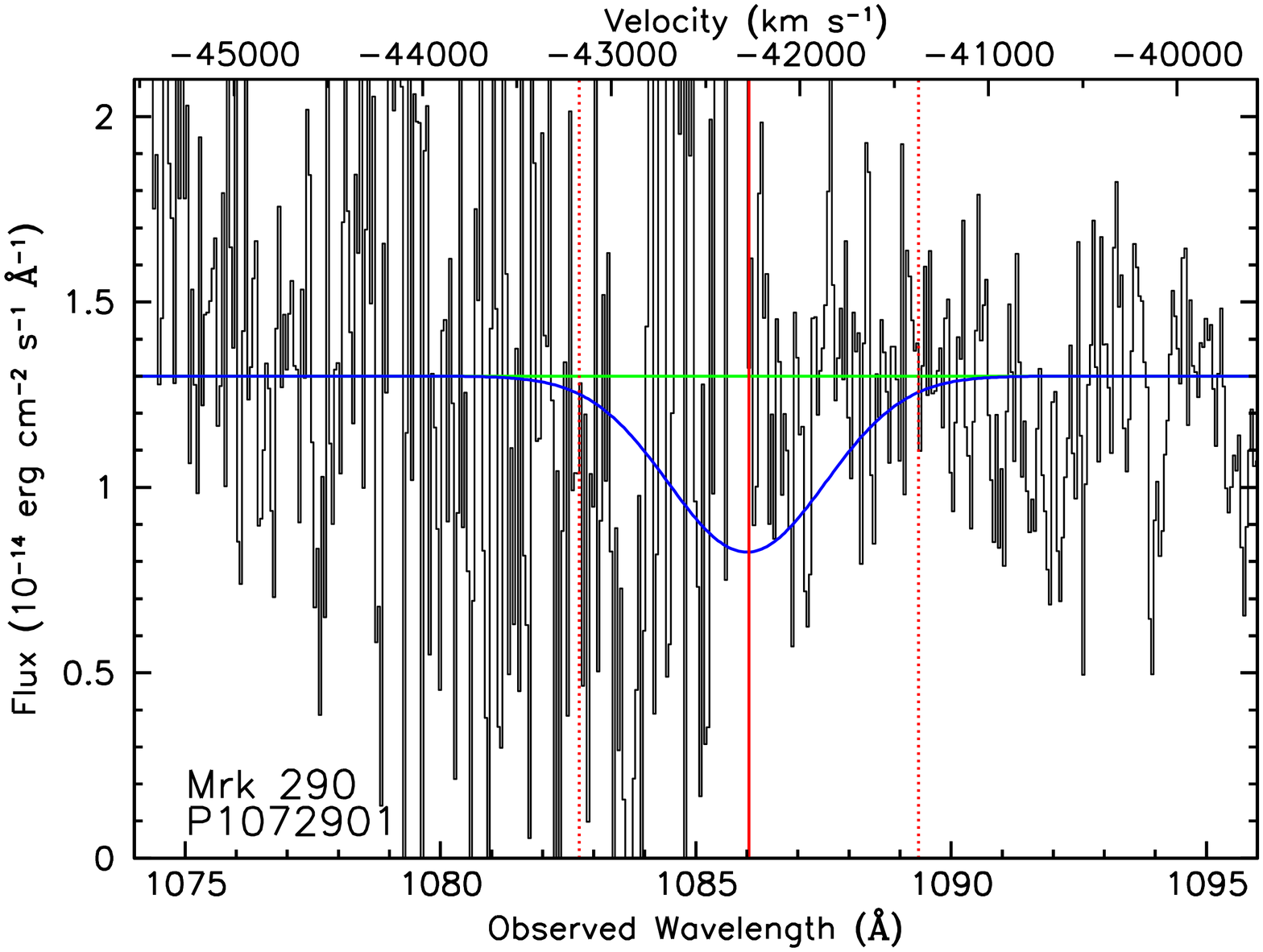}
    \caption{Mrk 290, FUSE observation P1072901}\label{fig:1p}
  \end{subfigure}%
  \begin{subfigure}[b]{.40\linewidth}
    \centering
    \includegraphics[angle=-90,width=.99\textwidth]{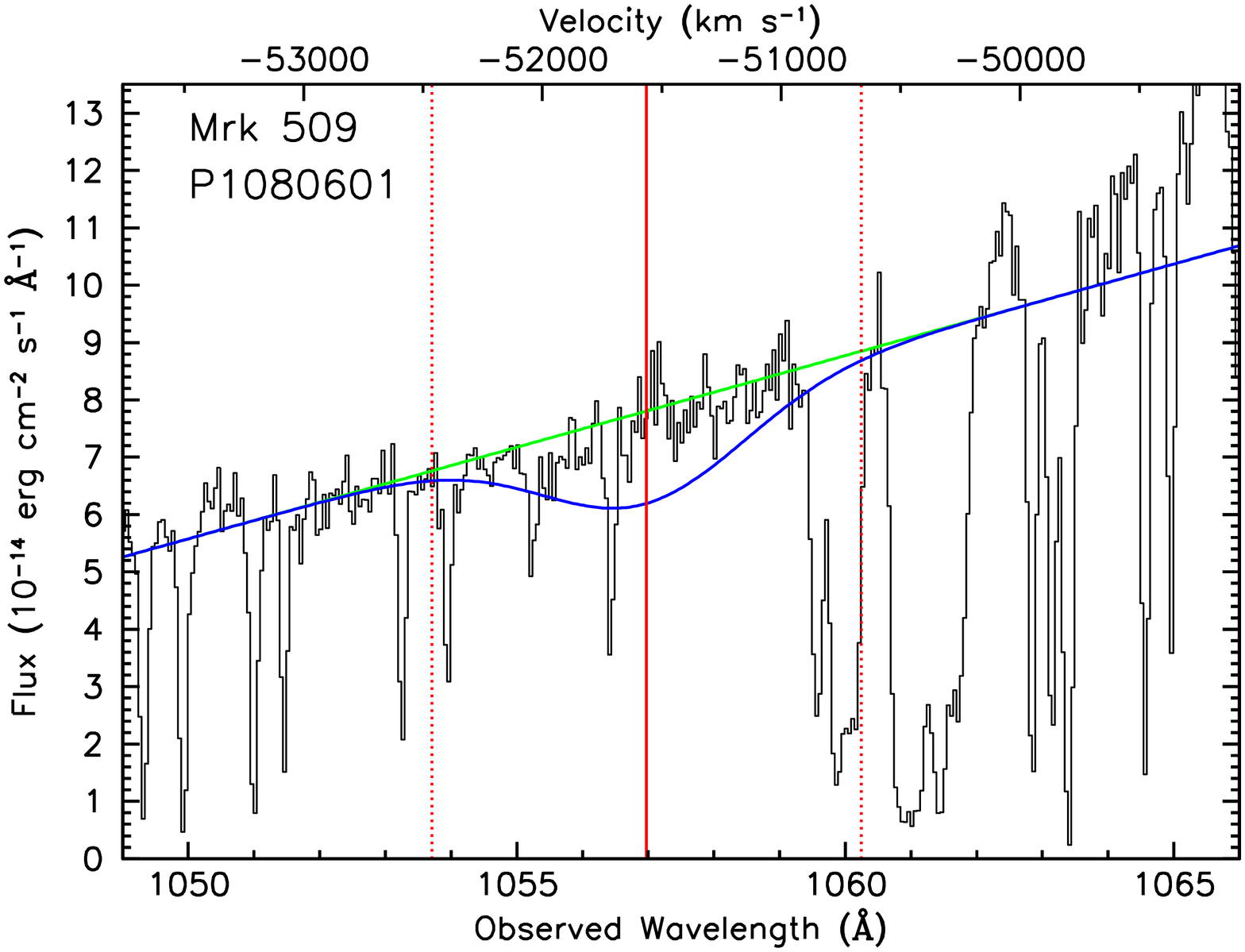}
    \caption{Mrk 509, FUSE observation P1080601}\label{fig:1q}
  \end{subfigure}%
  \begin{subfigure}[b]{.40\linewidth}
    \centering
    \includegraphics[angle=-90,width=.99\textwidth]{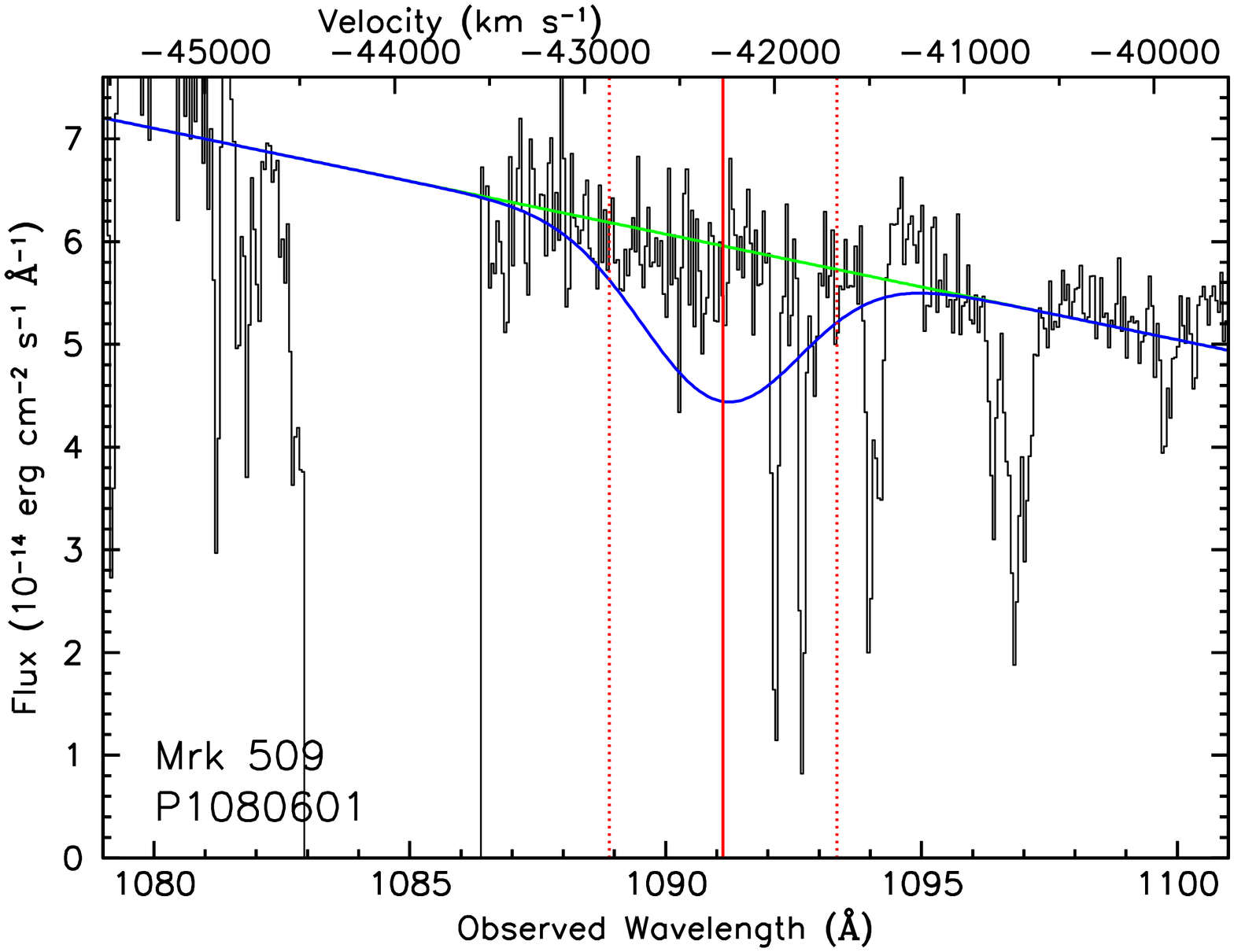}
    \caption{Mrk 509, FUSE observation P1080601}\label{fig:1r}
  \end{subfigure}%
  \begin{subfigure}[b]{.40\linewidth}
    \centering
    \includegraphics[angle=-90,width=.99\textwidth]{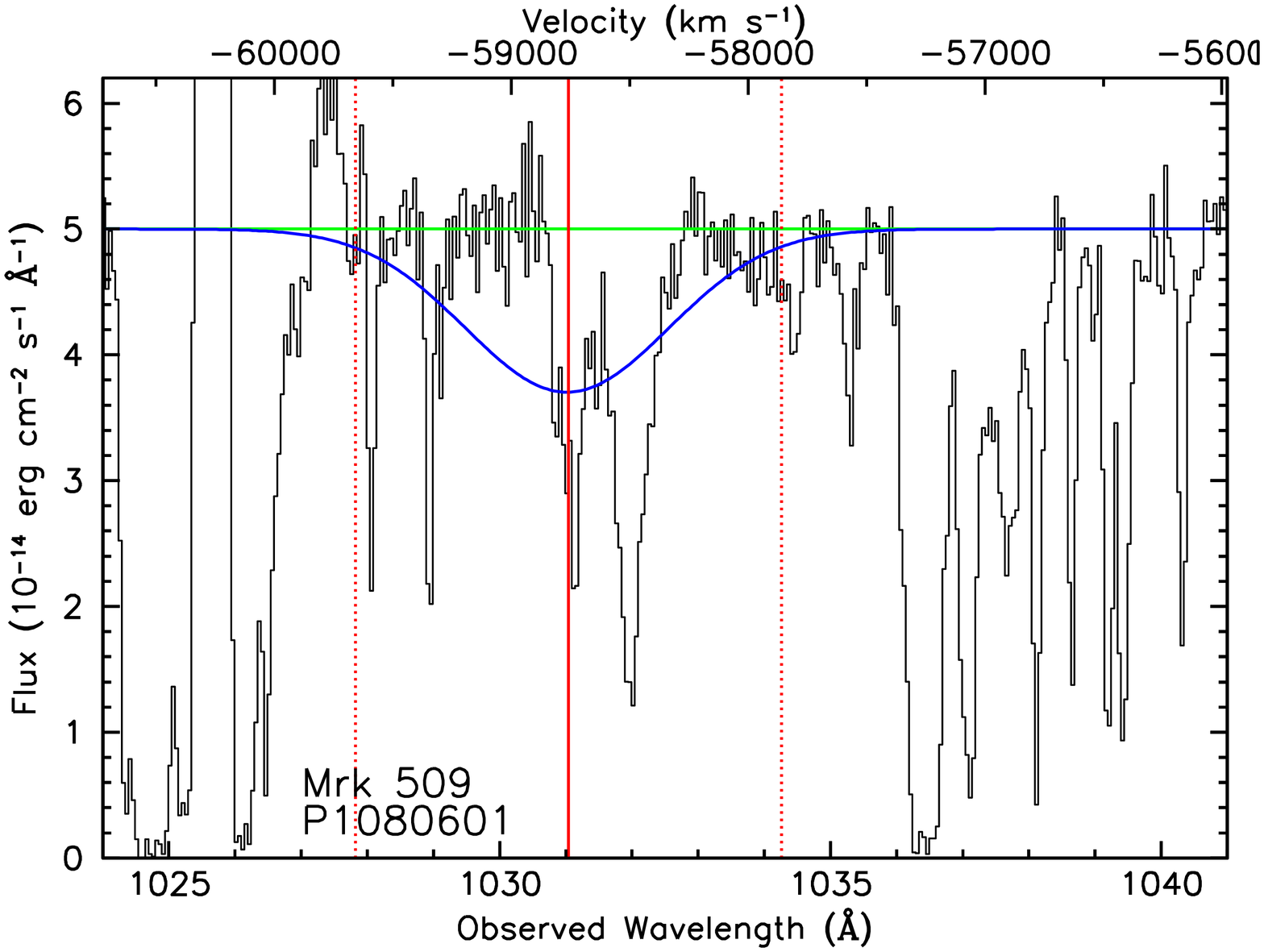}
    \caption{Mrk 509, FUSE observation P1080601}\label{fig:1s}
  \end{subfigure}%
  \begin{subfigure}[b]{.40\linewidth}
    \centering
    \includegraphics[angle=-90,width=.99\textwidth]{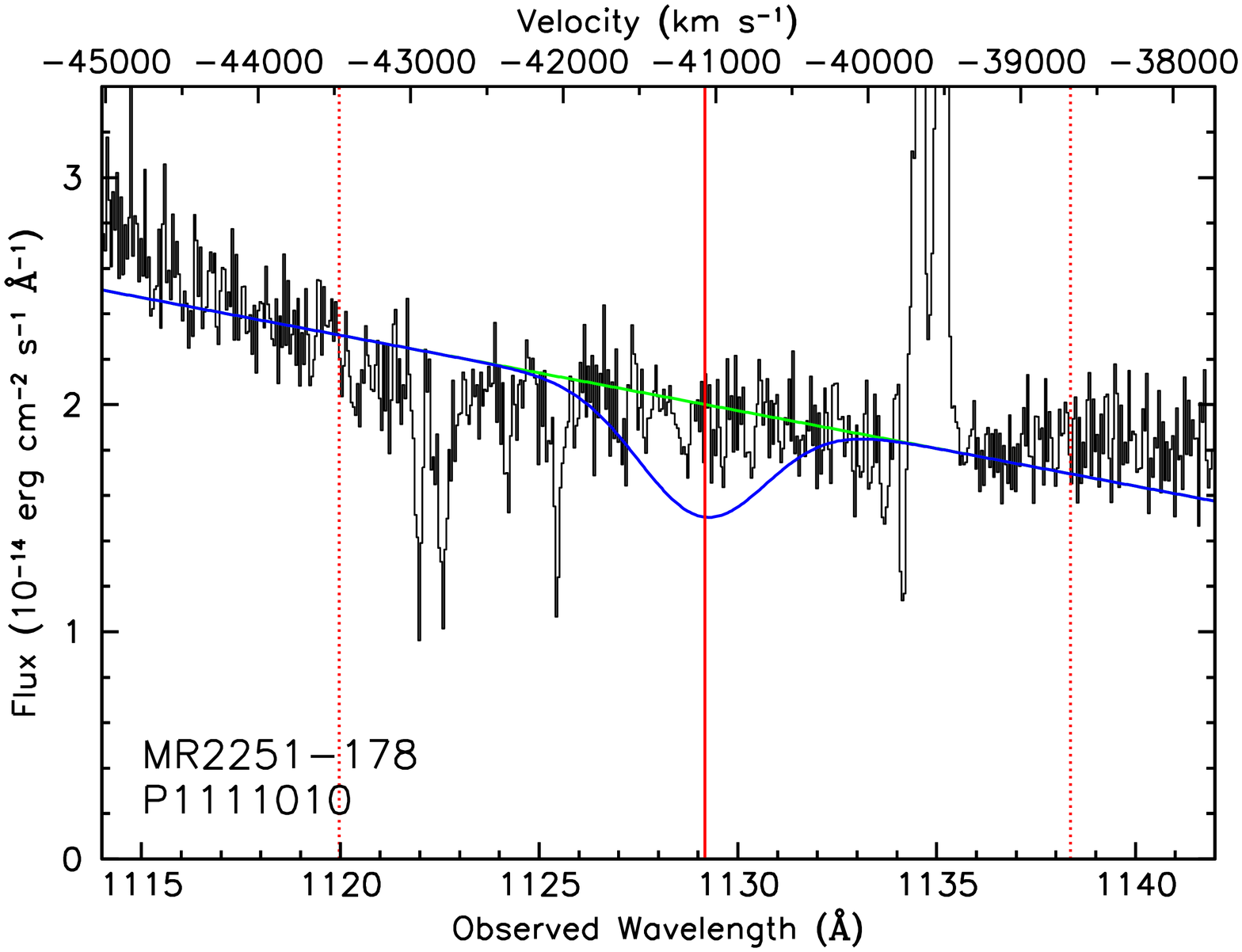}
    \caption{MR2251$-$178, FUSE observation P1111010}\label{fig:1t}
  \end{subfigure}\\%
  \caption{(cont.)}\label{FigFitsc}
  \addtocounter{figure}{-1} 
\end{figure*}

\end{document}